\documentclass[a4paper]{article}
\usepackage{graphicx,psfrag,amsmath,epsfig}
\usepackage{nameref}
\usepackage{journalhref}
\usepackage{hyperref}

\renewcommand{\theequation}{\arabic{section}.\arabic{equation}}
\renewcommand{\d}{\partial}
\newcommand{\dLR}{{\sss \overleftrightarrow{\textstyle \Delta}}}
\newcommand{\DLR}{{\sss \overleftrightarrow{\textstyle \mbold{D}}}}
\newcommand{\Int}{\!\int\!}
\newcommand{\dnu}{d\nu}
\newcommand{\mbold}[1]{{\bf{#1}}}
\newcommand{\mathbold}[1]{\mbox{\boldmath $\bf#1$}}

\newcommand{\pim}{\pi_{\scriptscriptstyle{-}}}
\newcommand{\pin}{\pi_{\scriptscriptstyle{0}}}
\newcommand{\Kp}{K_{\scriptscriptstyle{+}}}

\newcommand{\Kn}{K_{\scriptscriptstyle{0}}}
\newcommand{\Mpic}{M_{\pi^{\scriptscriptstyle{+}}}}
\newcommand{\Mpin}{M_{\pi^{\scriptscriptstyle{0}}}}
\newcommand{\MKc}{M_{\scriptscriptstyle{K^+}}}
\newcommand{\MKn}{M_{\scriptscriptstyle{K^0}}}
\newcommand{\w}[2]{\omega_#1(#2)}
\newcommand{\Lagr}{\mathcal{L}}
\newcommand{\Order}[1]{\mathcal{O}#1}
\newcommand{\order}[1]{{\it o}#1}
\newcommand{\nn}{\nonumber}
\newcommand{\sss}{\scriptscriptstyle}
\begin{document}
\begin{titlepage}
\vspace*{1cm}
\begin{center}
  {\LARGE\bf Spectra and decays of \boldmath{$\pi \pi$}\\ and \boldmath{$\pi K$} atoms}

\vspace{1.8cm}
J. Schweizer\\Institute for Theoretical Physics,\\
University of Bern,\\ 
Sidlerstr. 5, CH--3012 Bern, Switzerland\\
E-mail: schweizer@itp.unibe.ch

\vspace{0.3cm}
May 5, 2004
\vspace{0.6cm}

\parbox{\textwidth}{
We describe the spectra and decays of $\pi^+\pi^-$ and $\pi^\pm K^\mp$ atoms
within a non-relativistic effective field theory. The evaluations of the
energy shifts
and widths are performed at next-to-leading order in isospin symmetry
breaking. We provide general formulae for all S-states, and discuss the states
with angular momentum one in some detail. The prediction for the lifetime of the $\pi^\pm K^\mp$ atom in its
ground-state yields $\tau_{10}=(3.7\pm0.4)\cdot10^{-15}$s.}

\vspace{2cm}
\footnotesize{\begin{tabular}{ll}
{{\bf{Pacs}} number(s):}$\!\!\!\!$& 03.65.Ge, 03.65.Nk, 11.10.St, 12.39.Fe,
13.40.Ks \\
{\bf{Keywords:}}$\!\!\!\!$ & Hadronic atoms, Chiral perturbation theory,
Pion-pion scattering \\
& lengths, Pion-kaon scattering lengths, Non relativistic
effective \\
& Lagrangians, Isospin symmetry breaking, Electromagnetic corrections
\end{tabular}}
\end{center}
\end{titlepage}
\thispagestyle{empty}

\tableofcontents

\newpage
\setcounter{equation}{0}
\section{Introduction}
The DIRAC collaboration \cite{Adeva:1994xz} at CERN has measured the
lifetime of pionium in its ground-state, and the preliminary result yields
$\tau_{\pi,10}=[3.1^{+0.9}_{-0.7}({\rm stat})\pm 1({\rm syst})]\cdot10^{-15}$s
\cite{dirac}. A lifetime measurement of pionium at the $10\%$ level allows one to
determine the S-wave $\pi\pi$ scattering length difference $|a_0^0-a_0^2|$ at
$5\%$ accuracy. The measurement can then be compared with theoretical predictions for
the S-wave scattering lengths
\cite{Weinberg:1966kf,Colangelo:2000jc,Colangelo:2001df} and with the results coming from scattering experiments
\cite{Rosselet:1976pu}. Particularly exciting is the fact that
this enterprise subjects chiral perturbation theory to a very sensitive test
\cite{Knecht:1995tr}. 
New measurements are proposed for 
CERN PS, J-PARC and GSI \cite{Adeva:2000vb}. These experiments aim to measure
the lifetime of $\pi^+\pi^-$ and $\pi^\pm K^\mp$ atoms simultaneously.

In order to extract the scattering lengths from such future precision
measurements, the theoretical expressions for the energy shifts and decay
widths of the $\pi^+\pi^-$ and $\pi^\pm K^\mp$ atoms must be known to a precision
that matches the experimental accuracy. Nearly fifty years ago, Deser {\it et al.} \cite{Deser:1954vq} derived the
leading order formulae for the decay width and the energy shift in pionic
hydrogen. Similar relations exist for $\pi^+\pi^-$ and $\pi^\pm K^\mp$ atoms
\cite{Palfrey:kt,Bilenky:zd}, which decay due to the strong interactions into
$2\pi^0$ and $\pi^0 K^0$, respectively. Theoretical investigations on the
spectrum and the decay of pionium have been performed beyond leading order
in several settings. Potential
scattering has been used \cite{Trueman,Moor:ye,Minkowski} as well as field-theoretical
methods \cite{Efimov:1985fe,Belkov:1985xn,Volkov:ad,pipi,Lyubovitskij:1996mb,Jallouli:1997ux}. In particular, the
lifetime of pionium was studied by the use of the Bethe-Salpeter equation
\cite{Lyubovitskij:1996mb} and in the framework of the
quasipotential-constraint theory approach \cite{Jallouli:1997ux}. The width of the $\pi^+\pi^-$ atom has also been analyzed within a
non-relativistic effective field theory
\cite{Labelle:1998gh,Eiras:2000rh,Gasser:2001un}, which was originally
developed for bound states in QED by Caswell and Lepage
\cite{Caswell:1985ui}. The non-relativistic framework has proven to be a very efficient method to
evaluate bound state characteristics. It was further applied to the
ground-state of pionic hydrogen \cite{Lyubovitskij:2000kk,Gasser:2002am,Zemp} and very recently
to the energy-levels and decay widths of kaonic hydrogen
\cite{Meissner:2004jr}. Within the non-relativistic effective field theory the
isospin symmetry breaking corrections to the Deser-type formulae
can be evaluated systematically. In Refs. \cite{Labelle:1998gh,Eiras:2000rh,Gasser:2001un,Hammer:1999up,Gasser:1999vf} the lifetime of pionium was
evaluated at next-to-leading order in the isospin breaking parameters $\alpha\simeq1/137$ and $(m_u-m_d)^2$.

We presented in Ref. \cite{Schweizer:2004ir}, the results for the S-wave decay
widths and strong energy shifts of $\pi^+ \pi^-$ and $\pi^\pm K^\mp$ atoms at
next-to-leading order in isospin symmetry breaking. Further, for the lifetime as well as for the first two energy-level shifts, a numerical
analysis was carried out. The aim of this article is to provide the details
that have been omitted in Ref. \cite{Schweizer:2004ir}. Chiral perturbation
theory (ChPT)
allows one to relate the result for the width of the $\pi^\pm K^\mp$ atom to
the isospin odd $\pi K$ scattering lengths $a_0^-$, while the strong energy
shift is proportional to the sum of isospin even and odd scattering lengths
$a_0^++a_0^-$. The values for $a_0^+$ and $a_0^-$, used in the numerical
evaluation of the widths and strong energy shifts, stem from the recent
analysis of $\pi K$ scattering from Roy and Steiner type equations
\cite{Buettiker:2003pp}. Within
ChPT, the $\pi K$ scattering lengths have been worked out at one--loop accuracy
\cite{Bernard:1990kx,Roessl:1999iu,Nehme:2001wa}, and very recently even
the chiral expansion of the $\pi K$ scattering
amplitude at next-to-next-to-leading order became available \cite{Bijnens:2004bu}.
Particularly interesting is that the isospin even scattering lengths $a_0^+$
depends on the low--energy constant $L_6^r$
\cite{Gasser:1984gg}, and this coupling is related to the flavour dependence
of the quark condensate \cite{Descotes-Genon:1999uh}.

The paper is organized as follows: The general features of $\pi^+\pi^-$ and
$\pi^\pm K^\mp$ atoms are described in Section \ref{section: overview}. The non-relativistic effective field-theory
approach is illustrated in Section \ref{section: NR framework} by means of the
$\pi^-K^+$ atom. The discussion includes the Hamiltonian, the master equation,
and the matching to the relativistic $\pi K$ amplitudes. In
Sec. \ref{section: results}, we present the results for the decay widths and
strong energy shifts of the $\pi^\pm K^\mp$ and $\pi^+\pi^-$
atoms at next-to-leading order in isospin symmetry breaking. The role of
transverse photons is discussed in Section \ref{section:
  transverse}. Transverse photons do contribute to the electromagnetic part of
the energy shift. The pure QED contributions have been worked out a long time
ago, based on the Bethe-Salpeter equation \cite{Nandy:rj}, the
quasipotential approach \cite{Todorov:1970gr,Jallouli:1996bu} and an improved
Coulomb potential \cite{Austen:te}. We reproduced this result within the
non-relativistic framework.
We further estimate the contributions from transverse photons
to the decay width of the $\pi^- K^+$ atom and show that they vanish at
next-to-leading order in isospin symmetry breaking. The contributions
generated by the vacuum polarization of the electron \cite{Eiras:2000rh,Lyubovitskij:2000kk,vacpol} are
 discussed in Section \ref{section: vacpol}. Formally of higher order in
 $\alpha$, they are numerically not negligible. A numerical analysis of
the widths and the energy-level shifts is carried out in Section \ref{section:
  numerics} at $\Order(p^4,e^2p^2)$ in the chiral expansion. 

\setcounter{equation}{0}
\section[General features of \boldmath{$\pi\pi$} and \boldmath{$\pi K$} atoms]{General features of \boldmath{$\pi\pi$} and \boldmath{$\pi K$} atoms}
\label{section: overview}
In this section, we describe the general features of the systems that we are
going to study. The $\pi^+ \pi^-$ and $\pi^\pm K^\mp$ atoms are highly
non-relativistic, loosely bound systems, mainly formed by the Coulomb
interaction. The average momentum of the constituents in the c.m. frame lies in the MeV range. Further, their
decay widths $\sim 0.2$eV are much smaller than the binding energies $\sim
10^{3}$eV involved. The $\pi^+\pi^-$ atom in its ground-state decays predominantly into
a pair of two neutral pions, through the strong transition
$\pi^+\pi^-\rightarrow\pi^0\pi^0$. The decay width into two photons is
suppressed by the factor $4\cdot 10^{-3}$ \cite{Adeva:1994xz,Hammer:1999up}. For a detailed
discussion of the decay channels of pionium, we refer to \cite{Gasser:2001un}.
The decays of the $\pi^- K^+$ atom have to conserve strangeness. Apart from
the dominant S-wave decay channel into $\pi^0 K^0$, the only allowed decays are
therefore $K^0 + n_1\gamma+n_2e^+e^-$ and $\pi^0 K^0+n_1\gamma+n_2e^+e^-$,
where $n_1+n_2>0$. Here $n_1$ and $n_2$ denotes the number of photons and $e^+e^-$ pairs,
respectively. In the
relativistic theory, the odd intrinsic parity process $\pi^-K^+\rightarrow
K^0+2\gamma$ corresponds to a local interaction in the Wess-Zumino-Witten term
\cite{Wess:yu}, while the transition $\pi^-K^+\rightarrow K^0+\gamma$ occurs
not until $\Order{(p^6)}$
\cite{Bijnens:2001bb}.

The non-relativistic framework
\cite{Labelle:1998gh,Gasser:2001un,Caswell:1985ui,Lyubovitskij:2000kk} we are
going to apply, provides a systematic expansion in the isospin breaking
parameter $\delta$. In the case of pionium, we count $\alpha$ as well as
$(m_u-m_d)^2$ as small
quantities of order $\delta$. As for the $\pi^\pm K^\mp$ atom, both $\alpha$
and $m_u-m_d$ count as order $\delta$. The different
power counting schemes are due to the fact that in QCD, the chiral expansion
of the pion mass squared difference $\Delta_\pi = \Mpic^2-\Mpin^2$ is of second order in
$m_u-m_d$, while $\Delta_K = \MKc^2-\MKn^2$ is linear
in $m_u-m_d$.
At leading and next-to-leading order in isospin symmetry breaking, the
$\pi^+\pi^-$, ($\pi^-K^+$) atom decays into $\pi^0\pi^0$ ($\pi^0K^0$)
exclusively.  The leading order term for the width is of $\Order(\delta^{7/2})$,
isospin breaking corrections contribute at order $\delta^{9/2}$.
The results for the S-wave decay widths at
order $\delta^{9/2}$ are presented in Secs.
\ref{section: decayWidths} and \ref{section: pionium}. At order $\delta^5$,
also other decay channels contribute. In Section \ref{section: intStates}, we
estimate the order of the various decays.

The energy-level splittings of the $\pi^+ \pi^-$ and
$\pi^\pm K^\mp$ atoms are induced by both
electromagnetic and strong interactions. At order $\delta^3$, the energy shift contributions are exclusively due to strong interactions,
while at order $\delta^4$, both electromagnetic and strong interactions contribute. It is both conventional and convenient to split the energy shifts
into a strong and an electromagnetic part, according to\footnote{Note that this splitting cannot be understood literally, i.e. there are
contributions from strong interactions to $\Delta E^{\rm em}_{n0}$.}
\begin{equation}
  \Delta E_{nl} = \Delta E^{\rm h}_{nl}+\Delta E^{\rm em}_{nl}.
\label{DeltaEtot}
 \end{equation}
The expressions for the strong energy shift $\Delta E^{\rm h}_{nl}$ at
  next-to-leading order in isospin symmetry breaking are presented in Secs. \ref{section:
  energyShift} and \ref{section: pionium}. The electromagnetic part $\Delta
  E^{\rm em}_{nl}$ is discussed in Section \ref{section:
  emEnergy}. Another important correction is generated by the vacuum
  polarization of the electron. Formally,
  the vacuum polarization contributes to the energy shift at order $\delta^5$ and
  to the width at order $\delta^{11/2}$, but these corrections are amplified
  by powers of the ratio $\mu_+/m_e$. Here $\mu_+$ denotes the reduced mass of the bound
  system and $m_e$ the electron mass. The vacuum polarization contributions
  are discussed in Section \ref{section: vacpol}.

In what follows, we proceed systematically and discuss in detail the decays and
bound state spectra within the non-relativistic framework.
\setcounter{equation}{0}
\section{Non-relativistic framework}
\label{section: NR framework}
\subsection{Hamiltonian}
\label{section: hamiltonian}
The Hamiltonian consists of a infinite series of operators with increasing mass dimensions -
all operators allowed by gauge invariance, space rotation, parity and time reversal must be
included. However, in the evaluation of the decay width and the strong energy shift
at next-to-leading order in isospin symmetry breaking, only a few low dimensional
operators do actually contribute. For the $\pi^-K^+$ atom, the following Hamiltonian achieves the goal:
\begin{eqnarray}
\mbold{H} & =& \mbold{H}_0+\mbold{H}_{\sss \rm C}+\mbold{H}_{\sss \rm D}+\mbold{H}_{\sss \rm S},\nn\\
  \mbold{H}_{\rm \Gamma} &=& \int d^3 \mbold{x}\,\mathcal{H}_{\rm
  \Gamma}(0,\mbold{x}), \quad {\rm
  \Gamma} = {\rm 0,C,D,S},
\label{Hamilton}
\end{eqnarray}
with
\begin{eqnarray}
\mathcal{H}_0 &=& \sum_{i =
  \pm,0}\left\{\pi_i^\dagger\left(M_{\pi^i}-\frac{\Delta}{2M_{\pi^i}}\right)\pi_i+K_i^\dagger\left(M_{\sss
  K^i}-\frac{\Delta}{2M_{\sss
  K^i}}\right)K_i\right\},\nn\\
  \mathcal{H}_{\sss \rm D} &=& -\sum_{i =
  \pm,0}\left\{\pi_i^\dagger\left(\frac{\Delta^2}{8M_{\pi^i}^3}+\cdots\right)\pi_i+K_i^\dagger\left(\frac{\Delta^2}{8M_{\sss
  K^i}^3}+\cdots\right)K_i\right\},\nn\\
\mathcal{H}_{\sss \rm C} &=&-2\pi\alpha\left(\sum_\pm \pm
  \pi_\pm^\dagger\pi_\pm \pm K_\pm^\dagger
  K_\pm\right)\Delta^{-1}\left(\sum_\pm \pm \pi_\pm^\dagger\pi_\pm \pm
  K_\pm^\dagger K_\pm\right),\nn\\
\mathcal{H}_{\sss \rm S} &=&-C_1\pim^\dagger\Kp^\dagger\pim\Kp -C_2\left(
  \pim^\dagger\Kp^\dagger\pin\Kn+\textrm{h.c}\right)-
  C_3\pin^\dagger\Kn^\dagger\pin\Kn\nn\\
&& -C_4\left(\pim^\dagger\dLR\Kp^\dagger\pin\Kn+\textrm{h.c}\right)-C_5
  \left(\pim^\dagger\dLR\Kp^\dagger\pim\Kp+{\rm h.c}\right)\nn\\
&&-C_6(\pim^\dagger\pim)\Delta(\Kp^\dagger\Kp)-
  C_7\left(\nabla\pim^\dagger\Kp^\dagger\nabla\pin\Kn+\textrm{h.c}\right)+\cdots,
\end{eqnarray}
where $u\dLR v \doteq u \Delta v+ v\Delta u$.
We work in the c.m. system and thus omit terms proportional to the
c.m. momentum. The basis of operators with two space derivatives is
chosen such that none of them contributes to the S-wave decay width and
energy shift at the accuracy we are considering. For this reason, we
transformed the operator with two space derivatives on the neutral fields by
the use of the equations of motion,
  \begin{equation}
  \pim^\dagger \Kp^\dagger \pin \dLR \Kn =
  -4\mu_0(\Sigma_+-\Sigma_0)\pim^\dagger\Kp^\dagger\pin\Kn+\frac{\mu_0}{\mu_+}\pim^\dagger
  \dLR\Kp^\dagger\pin\Kn.
\label{eomidentity}
  \end{equation}
For the moment, we further neglect transverse photon contributions. To the
 accuracy we are working, they do not contribute to the decay width and to the strong energy-level
 shifts. However, transverse photons do contribute when we work out the
 electromagnetic energy-level shifts in Section \ref{section:
   emEnergy}. The non-relativistic Lagrangian in the presence of transverse photons is
 given in appendix \ref{appendix: Lagr}. 
The Hamiltonian in Eq. (\ref{Hamilton}) is built from the non-relativistic pion and kaon fields
 \begin{equation}
   \pi_i(0,\mbold{x}) = \int
   \dnu(\mbold{p})e^{i\mbold{p}\mbold{x}}\mbold{a}_i(\mbold{p}), \quad K_i(0,\mbold{x}) = \int
   \dnu(\mbold{p})e^{i\mbold{p}\mbold{x}}\mbold{b}_i(\mbold{p}), \quad i = \pm,0,
 \end{equation}
with $\dnu(\mbold{p}) \doteq d^3\mbold{p}/(2\pi)^3$ and 
\begin{eqnarray}
 \left[\mbold{a}_i(\mbold{p}),\mbold{a}_k^\dagger(\mbold{p}')\right] &=&
 (2\pi)^3\delta^3(\mbold{p}-\mbold{p}')\delta_{ik}, \nn\\
\left[\mbold{b}_i(\mbold{p}),\mbold{b}_k^\dagger(\mbold{p}')\right] &=&
 (2\pi)^3\delta^3(\mbold{p}-\mbold{p}')\delta_{ik}.
 \end{eqnarray}
The two particle states of zero total charge are defined by
 \begin{equation}
   \mid \!\mbold{p}_1,\mbold{p}_2\rangle_+ =
 \mbold{a}_-^\dagger(\mbold{p}_1)\mbold{b}_+^\dagger(\mbold{p}_2)\mid \! 0 \rangle,
 \quad \mid \!\mbold{p}_3,\mbold{p}_4\rangle_0 =
 \mbold{a}_0^\dagger(\mbold{p}_3)\mbold{b}_0^\dagger(\mbold{p}_4)\mid \! 0 \rangle,
 \end{equation}
and the total and reduced masses $\Sigma_i$ and $\mu_i$ respectively, read 
\begin{equation}
  \Sigma_i = M_{\pi^i}+M_{\sss K^i}, \quad
   \mu_i = \frac{M_{\pi^i}M_{\sss K^i}}{M_{\pi^i}+M_{\sss K^i}}, \quad i=+,0. 
\end{equation}
\subsection{Master equation}
\label{section: resolvents}
To evaluate the decay width and the strong energy shifts we make use of
resolvents. This technique, which was developed by Feshbach a
long time ago~\cite{Feshbach:ut}, has been discussed extensively in
Ref.~\cite{Gasser:2001un}.
To remove the center of mass momentum from the matrix elements of any
  operators $\mbold{R}$, we introduce the notation
\begin{eqnarray}
  \lefteqn{{}_a\langle \mbold{p}_1, \mbold{p}_2\!\mid \mathbold{R}(z) \mid\!\mbold{p}_3,\mbold{p}_4
  \rangle_b}\nn\\
 &=& (2\pi)^3 \delta^3(\mbold{p}_1+\mbold{p}_2-\mbold{p}_3-\mbold{p}_4)
{}_a(\mbold{p}_1, \mbold{p}_2\!\mid
  \mathbold{R}(z)\mid\!\mbold{p}_3,\mbold{p}_4)_b,
  \end{eqnarray}
where $a,b$ stand for $0,+$. Further, we have
\begin{equation}
  {}_a\!(\mbold{q}, -\mbold{q}\!\mid \mbold{R}(z)\mid \!
  \mbold{p}, -\mbold{p})_b
  \doteq {}_a\!(\mbold{q}\!\mid\mbold{R}(z)
  \mid\!\mbold{p})_b.
\end{equation}
The master formula to be solved is given by the following eigenvalue equation,
\begin{equation}
  z-E_n-\int \dnu(\mbold{P})\langle\Psi_{n0}, \mbold{P}\!\mid\mathbold{\bar{\tau}}(z)\mid\!\Psi_{n0},0\rangle = 0,
\label{masterformula}
\end{equation}
where $E_n = \Sigma_+-\alpha^2\mu_+/(2n^2)$
 denotes the $n$-th Coulomb energy and the unperturbed $n$-th eigenstate is given by 
\begin{equation}
  \mid\!\Psi_{n0},
\mbold{P}\rangle=\int \dnu(\mbold{q})\Psi_{n0}(\mbold{q})\mid\!\tfrac{\mu_+}{\MKc}\mbold{P}+\mbold{q},\tfrac{\mu_+}{\Mpic}\mbold{P}-\mbold{q}\rangle_+.
\end{equation}
Here $\Psi_{n0}(\mbold{q})$ stands for the Coulomb wave function of the bound $\pi^\pm
 K^\mp$ system in momentum space.
 The operator $\mathbold{\bar{\tau}}$, defined through
 \begin{equation}
  \mathbold{\bar{\tau}} 
= \mbold{V}+\mbold{V}\mbold{\bar{G}}^n_{\sss \rm C}\mathbold{\bar{\tau}}, \quad \mbold{V}=\mbold{H}_{\sss \rm D}+\mbold{H}_{\sss \rm S},
 \end{equation}
is regular in the vicinity of $E_n$.
The quantity $\mbold{\bar{G}}^n_{\sss \rm C}$ stands for the $n$-th energy eigenstate singularity removed Coulomb resolvent,
 \begin{equation}
   \mbold{\bar{G}}^n_{\sss \rm C} =
   \mbold{G}_{\sss \rm C}\bigg\{\mbold{1}-\int \dnu(\mbold{P})\mid\!\Psi_{n0}, \mbold{P}\rangle
   \langle \Psi_{n0}, \mbold{P}\!\mid\bigg\}, \quad \mbold{G}_{\sss \rm C} =
   \frac{1}{z-\mbold{H}_0-\mbold{H}_{\sss \rm C}}.
 \end{equation}
The master equation presents a compact form of the Rayleigh-Schr\"odinger perturbation theory. If we insert
$\mathbold{\bar{\tau}}$ iteratively into (\ref{masterformula}), the eigenvalue equation becomes 
\begin{equation}
  z = E_n -|\Psi_{n0}(\mbold{x}=0)|^2\left[C_1+C_2^2J_0(z)\right]+\cdots
\label{masterformulaLO}
\end{equation}
where $\Psi_{n0}(\mbold{x}=0)$ stands for the Coulomb wave function in
coordinate space
and $J_0$ denotes the loop integral in Eq. (\ref{Ji}). The function $J_0$ is
analytic in the complex $z$ plane, except for a cut on the real axis starting at $z=\Sigma_0$. The imaginary part of $J_0$ has the same sign as
${\rm \, Im}\,z$ throughout the cut $z$ plane, which does not allow
Eq. (\ref{masterformulaLO}) to have a solution on the first Riemann
sheet. However, if we analytically continue $J_0$ from the upper rim of the
cut to the second Riemann sheet, we find a solution at $z={\rm Re}\,z+i\,{\rm
Im}\,z$, with
\begin{equation}
  {\rm Re\,} z = E_n -\frac{\alpha^3\mu_+^3C_1}{\pi n^3}+\cdots, \quad {\rm Im\,} z =-\frac{\alpha^3\mu_+^3\mu_0}{2\pi^2n^3}C_2^2\sqrt{\rho_n}+\cdots,
\end{equation}
and $\rho_n = 2\mu_0\left(E_n-\Sigma_0\right)$.
In the following, we evaluate the S-wave decay width $\Gamma_{n0} = -2{\rm Im}\,z$ at order $\delta^{9/2}$ and the energy shift $\Delta
  E_{n0} = {\rm Re}\, z-E_n$ at order $\delta^4$.
We focus on the strong part of the energy shift only, the evaluation of
the electromagnetic energy shift is discussed in Section \ref{section:
  emEnergy}. As in Ref.~\cite{Gasser:2001un}, we reduce
  Eq.~(\ref{masterformula}) to a one-channel problem with an effective potential $\mbold{W}$,
\begin{equation}
 \varrho\mathbold{\bar{\tau}}\varrho =
  \varrho\mbold{W}\varrho+\varrho\mbold{W}\varrho\mathbold{\bar{G}}^n_{\sss \rm C}
  \varrho\mathbold{\bar{\tau}}\varrho,
\end{equation}
where
\begin{equation}
  \mbold{W} =
  \mbold{V}+\mbold{V}\varrho_0\mathbold{\bar{G}}_{\sss \rm C}^n\left\{\mbold{1}-\varrho_0\mbold{V}\varrho_0\mathbold{\bar{G}}_{\sss \rm C}^n\right\}^{-1}\varrho_0\mbold{V}.
\end{equation}
Here $\varrho$, ($\varrho_0$) denotes the charged (neutral) two-particle projector
\begin{equation}
 \varrho= \int \dnu (\mbold{p}_1) \dnu (\mbold{p}_2)\mid\!\mbold{p}_1, \mbold{p}_2\rangle_+ {}_+\!
 \langle\mbold{p}_1, \mbold{p}_2\!\mid, \quad \varrho_0 = \mbold{1}-\varrho.
\end{equation}
The matrix element of $\mbold{W}$ between charged states takes the
form\footnote{The delta function term contributes to the electromagnetic
  energy shift, see Eq. (\ref{emEnergy}).}:
\begin{eqnarray}
  {}_+\!(\mbold{q}\!\mid \mbold{W}(z) \mid\! \mbold{p})_+ &=&
  (2\pi)^3\delta^3(\mbold{q}-\mbold{p})\left[-\frac{\mbold{p}^4}{8}\left(\frac{1}{\Mpic^3}+\frac{1}{\MKc^3}\right)+\cdots\right]\nn\\
&& +w(z)+w_1(z)\mbold{p}^2+w_2(z)\mbold{q}^2+w_3(z)\mbold{p}\mbold{q}+\cdots\nn\\
\label{w}
\end{eqnarray}
 To the accuracy we are working, only the constant term $w(z)$ contributes to
 the decay width and strong energy shift. 
We get for the S-wave decay width of the $\pi^- K^+$ atom at order $\delta^{9/2}$,
\begin{equation}
  \Gamma_{n0} =-2\mid\!\Psi_{n0}(\mbold{x}=0)\!\mid^2{\rm Im}\,w(E_n)\left(1+2{\rm Re}\,
  w(E_n)\langle\mbold{\bar{g}}^n_{\sss \rm C}(E_n)\rangle\right)+\Order{(\delta^{5})},
\label{decaywidth}
\end{equation}
while the S-wave energy-level shifts due to strong interactions read at order $\delta^4$,
\begin{equation}
   \Delta E^{\rm h}_{n0} = {\mid\!\Psi_{n0}(\mbold{x}\!=\!0)\!\mid}^2{\rm Re}\, w(E_n)\left(
   1+{\rm Re}\,
   w(E_n)\langle\mbold{\bar{g}}^n_{\sss \rm C}(E_n)\rangle\right)+\Order{(\delta^{5})}.
\label{energyshift}
\end{equation}  
The quantity $\langle\mbold{\bar{g}}^n_{\sss \rm C}(E_n)\rangle$, given in appendix \ref{appendix: green function}, is related to
the integrated Schwinger Green function~\cite{Schwinger}. The real and imaginary part of $w(z)$ are given by
\begin{eqnarray} 
{\rm Re}\, w(E_n) &=& -C_1 + \frac{C_2^2 C_3\mu_0^2}{4\pi^2}\rho_n,\nn\\
 {\rm Im}\,w(E_n) &=& -\frac{\mu_0\sqrt{\rho_n}}{2\pi}C_2^2\left[1-\frac{C_3^2\mu_0^2\rho_n}{4\pi^2}+\frac{5\mu_0\rho_n}{8}\frac{\Mpin^3+\MKn^3}{\Mpin^3\MKn^3}\right].
\end{eqnarray}
The decay width (\ref{decaywidth}) and energy shift (\ref{energyshift}) still
depend on the effective couplings $C_i$, which have to be related to physical quantities.
\subsection{Matching procedure}
\label{section: matching}
We now determine the diverse couplings from matching the non-relativistic and
the relativistic amplitudes at threshold. With the effective Lagrangian in Eqs. (\ref{freeLagr}), (\ref{Lagr})
and (\ref{Lagr2}) we may calculate the non-relativistic $\pi^- K^+\rightarrow
\pi^0 K^0$ and $\pi^- K^+\rightarrow \pi^- K^+$ scattering amplitudes at
threshold at order $\delta$. Again, we may omit contributions from
transverse photons. The radiative corrections to the one-particle irreducible
amplitudes, generated by transverse photons, vanish at threshold at order $e^2$.

The coupling $C_3$ enters the decay width (\ref{decaywidth}) at order
$\delta^{9/2}$ and the energy shift (\ref{energyshift}) at
order $\delta^4$ and is therefore needed at $\Order{(\delta^0)}$ only. However, we have to determine both
  $C_1$ and $C_2$ at next-to-leading order in isospin symmetry breaking. The relativistic amplitudes are related to the non-relativistic ones through
\begin{equation}
  T^{lm;ik}_{\rm R}(\mbold{q};\mbold{p})
  = 4\left[\w{i}{\mbold{p}}\w{k}{\mbold{p}}\w{l}{\mbold{q}}\w{m}{\mbold{q}}\right]^{\frac{1}{2}}
  T^{lm;ik}_{\sss \rm NR}(\mbold{q};\mbold{p}),
\label{matching}
\end{equation}
where $\w{i}{\mbold{p}}=(M_i^2+\mbold{p}^2)^{1/2}$ and $\mbold{p}$, ($\mbold{q}$) denotes the incoming (outgoing) relative 3-momentum. In the isospin symmetry limit, only the lowest order of the non-relativistic
Lagrangian contributes at threshold and the effective couplings $C_1$, $C_2$
and $C_3$ yield
\begin{eqnarray}
  C_1 = \frac{2\pi}{\mu_+}\left(a_0^++a_0^-\right),\quad
  C_2 = -\frac{2\sqrt{2}\pi}{\mu_+}a_0^-,\quad
  C_3 = \frac{2\pi}{\mu_+}a_0^+,
\label{CI}
\end{eqnarray}
where $a_0^+$ and $a_0^-$ denote the isospin even and odd S-wave
scattering lengths, the notation used is specified in appendix \ref{appendix:
  relAmplitude}. The $\pi K$ scattering lengths are defined in QCD, at $m_u
  = m_d = \hat{m}$ and $M_\pi \doteq\Mpic$, $M_{\sss K}\doteq\MKc$.
\begin{figure}[ttbp]
\begin{center}
\leavevmode
\makebox{\includegraphics[height=1.9cm]{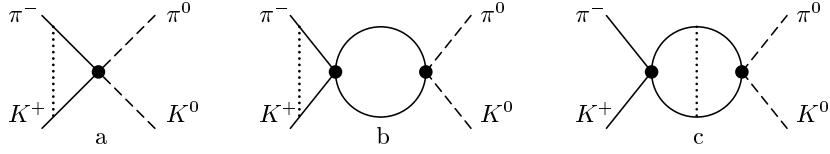}}
\caption{One-photon exchange diagrams for the $\pi^- K^+\rightarrow \pi^0 K^0$
  scattering amplitude. The dotted line denotes a Coulomb photon. The dots stand for the couplings $C_i$, $i=1,2$.}
\label{fig: Coulomb}
\end{center}
\end{figure}

To match the coupling $C_2$ including isospin symmetry breaking effects, we
calculate the real part of the non-relativistic
$\pi^-K^+ \rightarrow \pi^0K^0$ scattering matrix element in the vicinity of
the threshold at order $\delta$. In absence of virtual photons, the real part of the amplitude at threshold reads
\begin{equation}
  {\rm Re\,}T^{00;\pm}_{\sss \rm NR}(\mbold{q};\mbold{p})=C_2+C_2 C_3^2 J_0(\Sigma_+)^2+\cdots
\end{equation}
The ellipsis denotes terms which vanish a threshold or are of higher order in
  the parameter $\delta$. The one--loop integral $J_0$ is given in
  appendix \ref{appendix: integrals}. Bubbles with mass insertions and/or
  derivative couplings do not contribute at threshold at order $\delta$, since
  they contain additional factors of $\mbold{p}^2$ and/or $\Sigma_+-\Sigma_0$.

\begin{figure}[ttbp]
\begin{center}
\leavevmode
%\makebox{\epsfig{figure=Coulombel,height=5.5cm,bbllx=43,bblly=496,bburx=415,bbury=714,clip}}
\makebox{\includegraphics[height=5.3cm]{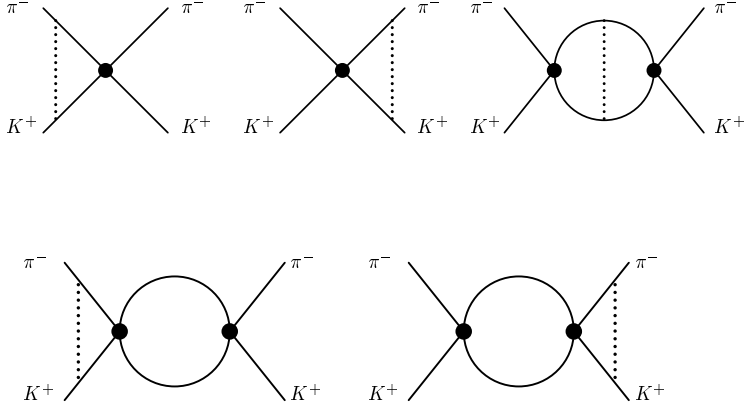}}
\caption{One-photon exchange diagrams for the truncated $\pi^- K^+\rightarrow \pi^- K^+$
  scattering amplitude. The dotted line denotes a Coulomb photon. The dot denotes the coupling $C_2$.}
\label{fig: Coulombel}
\end{center}
\end{figure}

We now include the Coulomb interaction. Feynman graphs, with a Coulomb
photon attached such that the heavy fields must propagate in time to
connect the two vertices, all vanish. This is because we may close the
integration contour over the zero-component of the loop momentum in the
half-plane where there is no singularity in the propagators. One example is the self-energy diagram, which
vanishes at order $\alpha$. As a result of this, there is no mass
renormalization in the non-relativistic theory and the mass parameters
$M_{\pi^i}$ and $M_{K^i}$, $i=0,+$ in the non-relativistic Lagrangian (\ref{Lagr}) stand for the physical meson masses. The amplitude at
threshold contains both infrared and ultraviolet singularities, coming from
the one-Coulomb photon exchange diagrams depicted in Figure \ref{fig: Coulomb}. Around threshold, we get for the one-Coulomb exchange diagrams, 
\begin{equation}
  T^{00;\pm}_{\sss \rm NR}(\mbold{q};\mbold{p})=-C_2V_{\sss \rm C}(\mbold{p}, P^0_{\rm thr})\left[1+C_1J_+(P^0_{\rm thr})\right]+C_1 C_2 B_{\sss \rm C}(P^0_{\rm thr})+\cdots,
\end{equation}
where $P^0_{\rm thr}=\Sigma_++\mbold{p}^2/(2\mu_+)$. The Coulomb vertex
  function $V_{\sss \rm C}$ in Figure \ref{fig:
  Coulomb}(a) and the two--loop integral
  $B_{\sss \rm C}$ in Fig. \ref{fig:
  Coulomb}(c) are given in appendix \ref{appendix:
  integrals}. The integral $J_+$, specified in Eq. (\ref{Jia}), has to be evaluated at
  $d\neq3$, because the vertex diagram generates an
  infrared singular Coulomb phase \cite{Yennie:ad} at threshold.
We split of this phase $\theta_c$, according to
\begin{eqnarray}
  T^{00;\pm}_{\sss \rm NR}(\mbold{q},\mbold{p}) &=&
  e^{i\alpha\theta_c}\hat{T}^{00;\pm}_{\sss \rm NR}(\mbold{q},\mbold{p}),\nn\\
\theta_c &=&
\frac{\mu_+}{|\mbold{p}|}\mu^{d-3}\left\{\frac{1}{d-3}-\frac{1}{2}\left[{\rm
      ln}4\pi+\Gamma'(1)\right]+{\rm ln}\frac{2|\mbold{p}|}{\mu}\right\},
\label{Coulombphase}
\end{eqnarray}
where $\mu$ denotes the running scale. The remainder $\hat{T}^{00;\pm}_{\sss \rm NR}$ is free of infrared
singularities at threshold, at order $\delta$. We find for the real part:
\begin{equation}
  {\rm Re}\,\hat{T}^{00;\pm}_{\sss \rm NR}(\mbold{q},\mbold{p}) = \frac{B_1}{|\mbold{p}|}+B_2
  {\rm ln}\frac{|\mbold{p}|}{\mu_+}+\frac{1}{N}{\rm Re}\,A^{00;\pm}_{{\rm
  thr}}+\Order{(\mbold{p})},
\label{ReTthr}
\end{equation}
with
\begin{equation}
  B_1 = C_2\frac{\alpha \pi \mu_+}{2}+\order{(\delta)}, \quad B_2 = -C_1
  C_2\frac{\alpha \mu_+^2}{\pi}+\order{(\delta)},
\end{equation}
and 
\begin{equation}
  N = 4\Mpic\MKc+\frac{\MKc-\Mpic}{\MKc+\Mpic}\left(\Delta_K-\Delta_\pi\right).
\end{equation}
\begin{figure}[ttbp]
\begin{center}
\leavevmode
\makebox{\includegraphics[height=2.2cm]{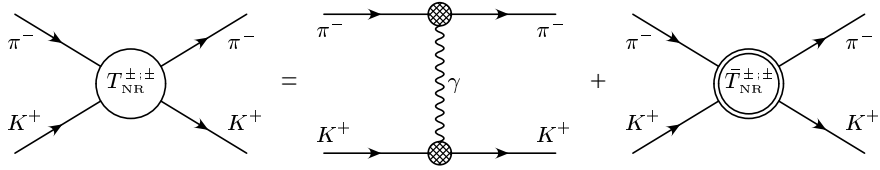}}
\caption{Non-relativistic $\pi^- K^+ \rightarrow \pi^- K^+$ scattering amplitude. The blob
  describes the vector form factor of the pion and
  kaon. $\bar{T}^{\pm;\pm}_{\sss \rm NR}$ denotes the truncated amplitude.}
\label{fig: TpiKel}
\end{center}
\end{figure}
At order $\delta$, the constant term in Eq.~(\ref{ReTthr}) reads
\begin{eqnarray}
  \frac{1}{N}{\rm Re}\,A^{00;\pm}_{\rm thr}&=& C_2\left\{1-
  C_3^2\frac{\mu_0^3(\Sigma_+-\Sigma_0)}{2\pi^2}+C_1\frac{\alpha\mu_+^2}{2\pi}\left[1-\Lambda(\mu)-{\rm
  ln}\frac{4\mu_+^2}{\mu^2}\right]\right\}\nn\\
&& +\order{(\delta)}.
\label{ReAthr}
\end{eqnarray}
The ultraviolet divergence $\Lambda(\mu)$, given in Eq. (\ref{Lambda}), stems
from the two--loop diagram $B_{\sss \rm C}$ and may be
absorbed in the renormalization of the coupling $C_2$,
\begin{equation}
   C_2^r(\mu) = C_2\left[1-
  \frac{\alpha\mu_+^2}{2\pi}C_1\Lambda(\mu)\right].
\end{equation}
We now determine the coupling constant $C_1$. At $\alpha = 0$, the real part
of the non-relativistic $\pi^-K^+ \rightarrow \pi^-K^+$ scattering amplitude reads at threshold,
\begin{equation}
  {\rm Re}\,T^{\pm;\pm}_{\sss \rm NR}(\mbold{p}, \mbold{p}) = C_1+C_3 C_2^2 J_0(\Sigma_+)^2+\cdots,
\end{equation}
where the ellipsis denotes contributions which vanish at threshold or are of
  $\order{(\delta})$. In the presence of virtual photons, we first have to subtract the
  one-photon exchange diagram from the full amplitude, as displayed in
  Fig. \ref{fig: TpiKel}. The coupling constant $C_1$ is now
  determined by the one-particle irreducible part of the amplitude. The
  truncated part $\bar{T}^{\pm;\pm}_{\sss \rm NR}$ again contains one-photon
  exchange diagrams as shown in Figure \ref{fig: Coulombel},
\begin{equation}
  \bar{T}^{\pm;\pm}_{\sss \rm
  NR}(\mbold{p}, \mbold{p}) =-2C_1 V_{\sss\rm C}(\mbold{p}, P^0_{\rm
  thr})\left[1+C_1J_+(P^0_{\rm thr})\right]+C_1^2B_{\sss\rm C}(P^0_{\rm thr})+\cdots
\end{equation}
All diagrams with a Coulomb photon exchange between an incoming and
an outgoing particle vanish, because the pions (kaons) must propagate in time
in order to connect the two vertices. Again the vertex function $V_{\sss\rm
  C}$ leads to an
infrared singular Coulomb phase at threshold,
\begin{equation}
  \bar{T}^{\pm;\pm}_{\sss \rm NR}(\mbold{p}, \mbold{p}) =
  e^{2i\alpha\theta_c}\hat{T}^{\pm;\pm}_{\sss \rm NR}(\mbold{p}, \mbold{p}),
\end{equation}
where $\hat{T}^{\pm;\pm}_{\sss \rm NR}$ is free of infrared singularities
at threshold at order $\delta$. 
Further, the real part of the infrared regular amplitude
$\hat{T}^{\pm;\pm}_{\sss \rm NR}$ is given by
\begin{equation}
{\rm Re}\,\hat{T}^{\pm;\pm}_{\sss \rm NR}(\mbold{p},\mbold{p}) = \frac{B_1'}{|\mbold{p}|}+B_2'
  {\rm ln}\frac{|\mbold{p}|}{\mu_+}+\frac{1}{4\Mpic\MKc}{\rm Re}\,A^{\pm;\pm}_{{\rm
  thr}}+\Order{(\mbold{p})},
\label{ReTthrc}
\end{equation}
with
\begin{equation}
  B_1' = C_1 \alpha \pi\mu_++\order{(\delta)}, \quad B_2' = -\frac{C_1^2 \alpha \mu_+^2}{\pi}+\order{(\delta)},
\end{equation}
and
\begin{eqnarray}
  \frac{1}{4\Mpic\MKc}{\rm Re}\,A^{\pm;\pm}_{{\rm
  thr}} &=& C_1\left\{1+\frac{C_1\alpha\mu_+^2}{2\pi}\left[1-\Lambda(\mu)-{\rm
  ln}\frac{4\mu_+^2}{\mu^2}\right]\right\}\nn\\
&&-\frac{C_2^{2}C_3\mu_0^3}{2\pi^2}(\Sigma_+-\Sigma_0)+\order{(\delta)}.
\label{ReAthrc}
\end{eqnarray}
Here, the ultraviolet pole term $\Lambda(\mu)$ in $B_{\sss \rm C}$ is
removed by renormalizing the coupling $C_1$, according to
\begin{equation} 
C_1^r(\mu) = C_1\left[1-
  \frac{\alpha\mu_+^2}{2\pi}C_1\Lambda(\mu)\right].
\end{equation}
The above renormalization of the low--energy couplings $C_1$ and $C_2$
eliminates at the same time the ultraviolet divergences contained in the
expressions for the decay width (\ref{decaywidth}) and the energy shift (\ref{energyshift}).
We assume that the relativistic $\pi K$ amplitudes at order $\delta$ have the same singularity structure as the non-relativistic amplitudes and 
use Eq.~(\ref{matching}) to match the non-relativistic expressions to the
relativistic ones. The calculations of the relativistic $\pi^-K^+\rightarrow
\pi^0K^0$ and $\pi^-K^+\rightarrow
\pi^-K^+$ scattering amplitudes
have been performed at $\Order{(p^4,e^2p^2)}$ in Refs.~\cite{Nehme:2001wa,Kubis:2001ij,Kubis}. Both the Coulomb phase and the
singular term $\sim \!{\rm ln}|\mbold{p}|$ are absent in the real part of the
amplitudes at this order of accuracy, they first occur at order $e^2p^4$. The quantity ${\rm Re}\,A^{00;\pm}_{\rm thr}$, (${\rm
  Re}\,A^{\pm;\pm}_{\rm thr}$) is determined by the constant term occurring in
the threshold expansion of the corresponding relativistic
amplitude. Further, the
relativistic calculations~\cite{Nehme:2001wa,Kubis:2001ij,Kubis}, contain the
same singular contribution $\sim \!1/|\mbold{p}|$ as the non-relativistic
amplitude in Eqs.~(\ref{ReTthr}) and (\ref{ReTthrc}). 

The results for the matching of the coupling constants $C_1^r(\mu)$ and $C_2^r(\mu)$
yield at next-to-leading order in isospin symmetry breaking,
\begin{eqnarray}
  C_2^r(\mu) &=&\frac{1}{N}{\rm Re}\,A^{00;\pm}_{\rm
  thr}-2\sqrt{2}\pi a_0^-\bigg[2(\Sigma_+-\Sigma_0)(a_0^+)^2\nn\\
&&+\alpha\left({\rm ln}\,\frac{4\mu_+^2}{\mu^2}-1\right)(a_0^+
  +a_0^-)\bigg],
\label{eq: matchingC2} 
\end{eqnarray}
and
\begin{eqnarray}
C_1^r(\mu) &=&\frac{1}{4\Mpic \MKc}{\rm Re}\,A^{\pm;\pm}_{\rm
  thr}+8\pi (\Sigma_+-\Sigma_0) a_0^+(a_0^-)^2\nn\\
&&+2\pi\alpha\left({\rm ln}\,\frac{4\mu_+^2}{\mu^2}-1\right)(a_0^+
  +a_0^-)^2.
\label{eq: matchingC1} 
\end{eqnarray}
\setcounter{equation}{0}
\section{Strong energy shift and width}
\label{section: results}
The matching relations in Section \ref{section: matching}, allow us to specify
the results for the decay width and the strong energy shift in terms of the
relativistic $\pi K$ scattering amplitudes at threshold. The expressions are
valid at next-to-leading order in isospin symmetry breaking, and to all orders
in the chiral expansion.

\subsection[S-wave decay width of the $\pi K$ atom]{S-wave decay width of the \boldmath{$\pi K$} atom}
\label{section: decayWidths}
The matching results in Eqs. (\ref{CI}) and (\ref{eq: matchingC2}) yield for the decay width at order $\delta^{9/2}$ in terms of the
 relativistic $\pi^- K^+ \rightarrow \pi^0 K^0$ amplitude at threshold,
\begin{eqnarray}
  \Gamma_{n0}&=&
  \frac{8\alpha^3\mu_+^2}{n^3}p^*_n\mathcal{A}^2\left(1+K_n\right),\nn\\
\mathcal{A} &=& -\frac{1}{8\sqrt{2}\pi}\frac{1}{\Sigma_+}{\rm
  Re}\,A^{00;\pm}_{\rm thr}+\order{(\delta)},
\label{Gammapi0K0}
\end{eqnarray}
where 
\begin{eqnarray}
 K_n &=& \frac{\Mpic
  \Delta_K+\MKc\Delta_\pi}{\Mpic+\MKc}(a^+_0)^2
\nn\\&&-4\alpha\mu_+(a^+_0+a^-_0)\left[\psi(n)-\psi(1)-\frac{1}{n}+{\rm
  ln}\frac{\alpha}{n}\right]+\order{(\delta)}.
\end{eqnarray}
and $\psi(n)=\Gamma'(n)/\Gamma(n)$. Aside from the kinematical factor $p^*_n$, the decay width is expanded in powers of $\alpha$ and
$m_u-m_d$. The outgoing relative 3-momentum
\begin{equation}
  p^*_n = \frac{1}{2 E_n}\lambda\left(E_n^2,\Mpin^2,\MKn^2\right)^{1/2},
\end{equation}
with $\lambda(x,y,z)=x^2+y^2+z^2-2x y-2x z-2y z$, is chosen such that the total
final state energy corresponds to the $n$-th energy eigenvalue of the
$\pi^- K^+$ atom. In the isospin limit, the $\pi^- K^+ \rightarrow \pi^0 K^0$
amplitude at threshold is determined by the isospin odd scattering length
$a^-_0$. In order to extract $a^-_0$ from the above result of the decay width,
we first have to subtract the isospin breaking contribution from the
amplitude. We expand the normalized amplitude in powers of the isospin breaking
parameter $\delta$,
\begin{equation}
  \mathcal{A}=a^-_0+\epsilon+\order{(\delta)}.
\label{eq: A}
\end{equation}
The isospin breaking corrections $\epsilon$ have been evaluated at
$\Order{(p^4,e^2p^2)}$ in Refs. \cite{Kubis:2001ij,Kubis}. See
also the comments in Section \ref{section: numerics}.
We may now rewrite the expression for the width in the following form:
\begin{equation}
  \Gamma_{n0} =
  \frac{8\alpha^3\mu_+^2}{n^3}p_n^*(a_0^-)^2\left(1+\delta_{{\sss K},
  n}\right)+\Order{(\delta^5)}, \quad\delta_{{\sss K},
  n} = \frac{2\epsilon}{a_0^-}+K_n.
\label{Gammanum}
\end{equation}
The corrections $\delta_{{\sss K},n}$ to the Deser-type formula have been
  worked out at order $\alpha$, $m_u-m_d$, $\alpha\hat{m}$ and
  $(m_u-m_d)\hat{m}$. The corrections $\delta_{{\sss K},n}$, where $n=1,2$ are given numerically in Table \ref{table: delta}.

For the P-wave decay width into $\pi^0 K^0$, the leading order term is 
proportional the square of the coupling $C_7$ and of order
$\delta^{13/2}$. After performing the matching, we get at leading order
\begin{eqnarray}
  \Gamma_{n1,\pi^0 K^0}
  &=&\frac{8(n^2-1)}{n^5}\alpha^5\mu_+^4{p^*}^3{(a_1^-)}^2,
\label{GammaP}
\end{eqnarray}
where $a_1^-$ denotes the P-wave scattering length.
\subsection[Strong energy shift of the $\pi K$ atom]{Strong energy shift of the \boldmath{$\pi K$} atom}
\label{section: energyShift}
With the matching results in Eqs. (\ref{CI}) and (\ref{eq: matchingC1}), we may
specify the S-wave energy shifts at order $\delta^4$, in terms of the
 relativistic truncated $\pi^- K^+ \rightarrow \pi^- K^+$ amplitude at threshold,
\begin{equation}
  \Delta E^{\rm h}_{n0} =
  -\frac{2\alpha^3\mu_+^2}{n^3}\mathcal{A}'\left(1+K_n'\right),\quad
\mathcal{A}'=\frac{1}{8\pi\Sigma_+}{\rm Re}\,A^{\pm;\pm}_{{\rm
  thr}}+\order{(\delta)},
\label{DeltaE}
\end{equation}
with
\begin{equation}
K_n' = -2\alpha\mu_+(a_0^++a_0^-)\left[\psi(n)-\psi(1)-\frac{1}{n}+{\rm
  ln}\frac{\alpha}{n}\right]+\order{(\delta)}.
\end{equation}
For the ground-state, the result agrees with the one
obtained for the strong energy shift in pionic hydrogen
\cite{Lyubovitskij:2000kk}, if we replace $\mu_+$ with the reduced
mass of the $\pi^- p$ atom and ${\rm Re}\,A^{\pm;\pm}_{{\rm
  thr}}$ with the regular part of the $\pi^- p$ amplitude at
threshold. 

In the isospin limit, the normalized amplitude $\mathcal{A}'$ reduces to the
sum of the isospin even and odd scattering lengths $a^+_0+a^-_0$. Again, we expand $\mathcal{A}'$ in powers of $\alpha$ and $m_u-m_d$,
\begin{equation}
  \mathcal{A}'=a^+_0+a^-_0+\epsilon'+\order{(\delta)}.
\label{eq: Ap}
\end{equation}
The corrections $\epsilon'$ have been evaluated at $\Order{(p^4,e^2p^2)}$ in
Refs. \cite{Nehme:2001wa,Kubis}. See
also the comments in Section \ref{section: numerics}.  The isospin breaking corrections to the Deser-type formula read at
next-to-leading order:
\begin{equation}
  \Delta E_{n0}^{\rm h} =
  -\frac{2\alpha^3\mu_+^2}{n^3}(a_0^++a_0^-)\left(1+\delta_{{\sss K},
  n}'\right)+\Order{(\delta^5)}, \quad \delta_{{\sss K},n}' = \frac{\epsilon'}{a^+_0+a^-_0}+ K_n',
\label{DeltaEnum}
\end{equation}
where $\delta_{{\sss K},n}'$ has been worked out at order $\alpha$, $m_u-m_d$,
$\alpha\hat{m}$ and $(m_u-m_d)\hat{m}$ in the chiral expansion. For the first
two energy-levels, the numerical
values for $\delta_{{\sss K},n}'$ are given in Table \ref{table:
  delta}.
What concerns the energy splittings for $l=1$ and $n\geq2$, the leading order
contribution is given by
\begin{equation}
  \Delta E^{\rm h}_{n1} = -2C_6
  \nabla\Psi_{n1}^*(\mbold{x}=0)\nabla\Psi_{n1}(\mbold{x}=0).
\label{DeltaEl}
\end{equation}
Here $\Psi_{n1}$ denotes the Coulomb wave function with angular momentum $l=1$. The low energy coupling constant $C_6$ is determined through the $l=1$
partial wave contribution to the relativistic $\pi^- K^+ \rightarrow\pi^- K^+$
scattering amplitude and we find for the energy shift,
\begin{equation}
  \Delta E^{\rm h}_{n1} =
  -\frac{2(n^2-1)}{n^5}\alpha^5\mu_+^4\left(a_1^++a_1^-\right).
\label{DeltaEP}
\end{equation}
The result is proportional to the combination $a_1^++a_1^-$ of
P-wave scattering lengths and suppressed by a factor of $\alpha^5$.

\subsection{Pionium}
\label{section: pionium}
For pionium, we adopt the convention used in Refs. \cite{Gasser:2001un,
  Gasser:1999vf} and count $\alpha$ and $(m_u-m_d)^2$ as small isospin
  breaking parameters of
  order $\delta$. The decay width and energy shifts of the $\pi^+\pi^-$ atom can be obtained from the formulae (\ref{decaywidth}), (\ref{energyshift}) and
(\ref{DeltaEl}) through the following substitutions of the masses $\MKc\rightarrow \Mpic, \MKn\rightarrow \Mpin$ and coupling constants\footnote{The $c_i$ are the low--energy constants occurring
  in Refs.~\cite{Gasser:2001un,Gasser:1999vf}.},
\begin{equation}
C_1\rightarrow c_1, \quad C_2\rightarrow \sqrt{2}(c_2-2c_4 \Delta_\pi), \quad
  C_3\rightarrow 2c_3, \quad C_6\rightarrow c_6.
\end{equation}
The factor 2 in substituting $C_3$ comes from the different
normalization of the $\pi^0\pi^0$ state
$|\mbold{p}_3,\mbold{p}_4\rangle_0=a_0^\dagger(\mbold{p}_3)a_0^\dagger(\mbold{p}_4)|0\rangle$.
For the coupling constant $C_2$, the substitution is non-trivial because our
basis of operators with two space derivatives differs from the one used in
Refs. \cite{Gasser:2001un,Gasser:1999vf}. See also the comment in Section
\ref{section: hamiltonian}. The result for the S-wave decay width of pionium reads at order $\delta^{9/2}$,
\begin{eqnarray}
  \Gamma_{\pi, n0} =
  \frac{2}{9n^3}\alpha^3p^*_{\pi, n}\mathcal{A}_\pi^2\left(1+K_{\pi,
  n}\right),\quad \mathcal{A}_\pi = a_0^0-a_0^2+\epsilon_\pi+\order{(\delta)},
\label{Gammapi}
\end{eqnarray}
where
\begin{eqnarray}
  K_{\pi, n} &=&
  \frac{\kappa}{9}\left(a_0^0+2a_0^2\right)^2-\frac{2\alpha}{3}\left(2a_0^0+a_0^2\right)\left[\psi(n)-\psi(1)-\frac{1}{n}+{\rm
  ln}\frac{\alpha}{n}\right]+\order{(\delta)},\nn\\
p^*_{\pi, n}&=&\left(\Delta_\pi-\frac{\alpha^2}{4n^2}\Mpic^2\right)^{1/2},
\end{eqnarray}
and $\kappa =\Mpic^2/\Mpin^2-1$. The quantity $\mathcal{A}_\pi$ is defined
  as in Refs. \cite{Gasser:2001un,Gasser:1999vf}. The isospin
  symmetry breaking contributions $\epsilon_\pi$ have been evaluated at $\Order{(p^4, e^2p^2)}$ in
  Refs. \cite{Gasser:2001un,Gasser:1999vf,Knecht:1997jw}. The 
 corrections $\epsilon_\pi$ are of the order
  of $\alpha$ and $\alpha \hat{m}$. This is due to the fact that in the
  $\pi^+\pi^-\rightarrow\pi^0\pi^0$ scattering amplitude at threshold, the quark mass difference shows up at
  order $(m_u-m_d)^2\hat{m}$ only. For the decay width
  of the ground-state at order $\delta^{9/2}$, we reproduce the result obtained in
  Refs. \cite{Jallouli:1997ux,Gasser:2001un,Gasser:1999vf}. 
Again we may rewrite the formula for the width:
\begin{equation}
 \Gamma_{\pi, n0} = \frac{2\alpha^3}{9n^3}p_{\pi,
 n}^*(a_0^0-a_0^2)^2\left(1+\delta_{\pi, n}\right)+\Order{(\delta^5)}, \quad \delta_{\pi,
 n}=\frac{2\epsilon_{\pi}}{a_0^0-a_0^2}+K_{\pi, n}.
\label{Gammanumpi}
\end{equation}
The parameter $\delta_{\pi,n}$ contains the isospin breaking corrections to
the Deser-type formula at next-to-leading order. The numerical values for
$\delta_{\pi,n}$, with $n=1,2$  are listed in Table \ref{table: delta}. The decay width
  of the P-states into a pair of two neutral pions is forbidden by $C$ invariance. 

The strong energy shift of the $\pi^+\pi^-$ atom at order $\delta^4$ yields
\begin{eqnarray}
  \Delta E_{\pi, n0}^{\rm h} &=&
  -\frac{\alpha^3\Mpic}{n^3}\mathcal{A}_\pi'\left(1+K_{\pi, n}'\right),\nn\\
\mathcal{A}_\pi'&=&\frac{1}{6}\left(2a_0^0+a_0^2\right)+\epsilon_\pi',\nn\\
K_{\pi, n}' &=& -\frac{\alpha}{3}\left(2a_0^0+a_0^2\right)\left[\psi(n)-\psi(1)-\frac{1}{n}+{\rm
  ln}\frac{\alpha}{n}\right]+\order{(\delta)},
\label{DeltaEpi}
\end{eqnarray}
where $\mathcal{A}_\pi'$ is defined analogously to the quantity $\mathcal{A}'$
 discussed in Section \ref{section: energyShift}. 
 The isospin symmetry breaking contributions $\epsilon_\pi'$ have been
 calculated at $\Order{(p^4,e^2p^2)}$ in
 Refs. \cite{Meissner:1997fa,Knecht:2002gz}. Again the corrections
 $\epsilon_\pi'$ are of order $\alpha$ and $\alpha\hat{m}$. At order $\delta^4$, the Deser-type formula is
 changed by isospin breaking corrections, according to
\begin{equation}
\Delta E_{\pi, n0}^{\rm h}
  =-\frac{\alpha^3\Mpic}{6n^3}(2a_0^0+a_0^2)\left(1+\delta_{\pi, n}'\right)+\Order{(\delta^5)},
  \quad \delta_{\pi, n}' = \frac{6\epsilon_\pi'}{2a_0^0+a_0^2}+ K_{\pi,n}'.
\label{DeltaEnumpi}
\end{equation}
For the first two energy-levels, the numerical values for $\delta_{\pi,n}'$ are given in Table \ref{table:
  delta}. Finally, the leading order contribution to the strong energy-level shift for $l=1$
and $n \geq2$ reads 
  \begin{equation}
  \Delta E^{\rm h}_{\pi, {n1}} = -\frac{(n^2-1)}{8n^5}\alpha^5\Mpic^3 a_1^1,
\label{DeltaEPpi}
\end{equation}
here $a_1^1$ denotes the P-wave $\pi \pi$ scattering length.
\setcounter{equation}{0}
\section{Transverse photons}
\label{section: transverse}
We now concentrate on the contributions coming from transverse
photons. At order $\delta^4$, the energy-level shifts in $\pi^+\pi^-$ and
$\pi^\pm K^\mp$ atoms contain apart from the strong energy shift also
an electromagnetic contribution as well as finite size effects due
to the electromagnetic form-factors of the pion and kaon. We further
discuss the contributions from transverse photons to the decay width of the
$\pi^- K^+$ atom and show that they do not contribute at order
$\delta^{9/2}$. For pionium, the various higher order decay channels have been discussed in Ref. \cite{Gasser:2001un}. 
\subsection{Electromagnetic energy-level shifts}
\label{section: emEnergy}
As mentioned in Section \ref{section: overview}, we split the total energy
shift $\Delta E_{nl}$ in Eq. (\ref{DeltaEtot}) into the strong part displayed in
Eq. (\ref{DeltaE}) and an electromagnetic contribution $\Delta E^{\rm em}_{nl}$.
The electromagnetic part is of order $\alpha^4$ and contains both pure QED
corrections as well as finite size effects due to the charge radii of the pion
and kaon, see appendix \ref{appendix: Lagr}. The energy
shift contributions due to pure QED have been evaluated by the use of the
Bethe-Salpeter equation \cite{Nandy:rj}, the quasipotential approach
\cite{Todorov:1970gr,Jallouli:1996bu} and an improved Coulomb potential \cite{Austen:te}. 
Nevertheless, we find it useful to provide the calculation within the
non-relativistic framework.

Again, we start with the master equation (\ref{masterformula}), but instead
of the effective potential $\mbold{W}$, we consider
the operator $\mathbold{\bar{\tau}}$ in the second iterative approximation,
\begin{equation}
  \mathbold{\bar{\tau}} = \mbold{V}+\mbold{V}\mbold{\bar{G}}^n_{\sss \rm C}\mbold{V}+\Order{(\mbold{V}^3)}.
\end{equation}
The non-relativistic Lagrangian including transverse photons (\ref{freeLagr}),
(\ref{Lagr}) and (\ref{Lagr2})
gives rise to the following perturbation,
\begin{eqnarray}
  \mbold{V} &=& \mbold{H}_{\sss \rm D}+ \mbold{H}_{\sss \rm S}+e
  \mbold{H}_\gamma+e^2\lambda \mbold{H}_{\lambda},\nn\\
\mathcal{H}_\gamma &=& i\mbold{A}\left[\frac{1}{\Mpic}\pim^\dagger\nabla
  \pim-\frac{1}{\MKc}\Kp^\dagger \nabla\Kp\right],\nn\\
\mathcal{H}_{\lambda} &=& \pim^\dagger \Kp^\dagger\pim \Kp.
\end{eqnarray}
The photon field $\mbold{A}$ is given by
\begin{equation}
  \mbold{A}(0,\mbold{x}) = \int \frac{d^3\mbold{k}}{(2\pi)^32k^0} \sum_{\lambda=1,2}
  \left[\mathbold{\epsilon}(k,\lambda) \,a_\gamma(k,\lambda)e^{i\mbold{k}\mbold{x}}+\mathbold{\epsilon}^*(k,\lambda) a_\gamma^\dagger(k,\lambda)
  e^{-i\mbold{k}\mbold{x}}\right],
\end{equation}
where $k^0 \doteq |\mbold{k}|$ and $\mathbold{\epsilon}(k,\lambda)$ denote the
transversal polarization vectors.
%$\mathbold{\epsilon}(k,\lambda)$ fulfill the completeness relation
% \begin{equation}
%   \sum_{\lambda=1,2} \epsilon^i(k,\lambda)\epsilon^{* j}(k,\lambda) = \delta^{ij}-\frac{k^i k^j}{\mbold{k}^2}.
% \end{equation}
The operator $a_\gamma$ satisfies the commutation relation,
\begin{equation}
   \left[a_\gamma(k,\lambda),a^\dagger_\gamma(k',\lambda')\right] = 2k^0(2\pi)^3\delta^3(\mbold{k}-\mbold{k}')\delta_{\lambda\lambda'},
\end{equation}
and the one-photon states read
\begin{equation}
  |\mbold{k},\lambda\rangle = a^\dagger_\gamma(k,\lambda)|0\rangle.
\end{equation}
\begin{figure}[t]
\begin{center}
\leavevmode
\makebox{\includegraphics[height=2.8cm]{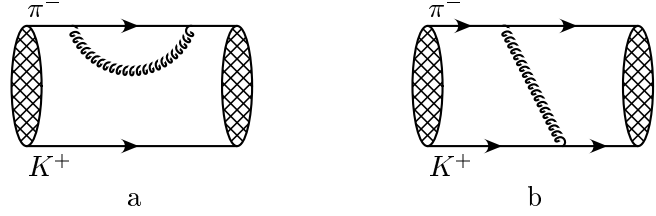}}
\caption{Self-energy (a) and one-photon exchange (b) to the electromagnetic
  energy shift. The twisted lines denote transverse photons.}
\label{fig: transverse}
\end{center}
\end{figure}
The electromagnetic contributions to the energy-level shifts consists of
\begin{eqnarray}
  \Delta E^{\rm em}_{nl} &=&
  -\frac{1}{8\mu_+^3}\left(1-\frac{3\mu_+}{\Sigma_+}\right)\int d^3\mbold{r}\,\Psi^*_{nl}(\mbold{r})\Delta^2\Psi_{nl}(\mbold{r})+e^2\lambda|\Psi_{n0}
  (\mbold{x}=0)|^2\nn\\
&&-\int
  \dnu(\mbold{p})|\Psi_{nl}(\mbold{p})|^2\left[\hat{\Sigma}_{\pi}(\Omega_n,\mbold{p})+\hat{\Sigma}_{\sss
  K}(\Omega_n,\mbold{p})\right]\nn\\
&& -\frac{e^2}{\Mpic
    \MKc}\int
  \dnu(\mbold{p})\dnu(\mbold{p}')\Psi_{nl}^*(\mbold{p})G_\gamma(\mbold{p},\mbold{p}')\Psi_{nl}(\mbold{p}'),
\label{emEnergy}
\end{eqnarray}
with $\Omega_n =\Sigma_++\mbold{p}^2/(2\mu_+)-E_n$ and $\Psi_{nl}$ denotes the
  Coulomb wave function for arbitrary $n$ and $l$. Here, the first term contains the mass insertions $\mbold{H}_{\sss \rm
  D}$, the second describes the finite size effects due to $\mbold{H}_{\lambda}$,
  while the last two terms come from the self-energy and one-photon exchange
  diagrams depicted in Figure \ref{fig: transverse} (a) and (b).

To avoid contributions from hard photon momenta, we
  use the threshold expansion \cite{threxp1,threxp2} to evaluate the
  self-energy contributions. This procedure is outlined in appendix
  \ref{appendix: integrals} and the threshold expanded self-energy
  $\hat{\Sigma}_{h}$, where $h=\pi, K$, is specified in
  Eq. (\ref{thrselfenergy}). As can be read off from the wave function in
  momentum space, the relative 3-momentum $\mbold{p}$ is of order
  $\delta$. Hence the quantities $\hat{\Sigma}_{\pi}(\Omega_n,\mbold{p})$ and
  $\hat{\Sigma}_{\sss K}(\Omega_n,\mbold{p})$ count as order $\delta^5$
  and are beyond the accuracy of our calculation.

What remains to be calculated is the one-photon exchange contribution\footnote{We thank A. Rusetsky for a very useful
  communication concerning technical aspects of the calculation.}. The
integrand
  \begin{eqnarray}
    G_\gamma(\mbold{p},\mbold{p}') &=&\frac{1}{|\mbold{p}-\mbold{p}'|}\left[\frac{\alpha^2\mu_+}{2n^2}+\frac{\mbold{p}^2}{2\Mpic}+\frac{\mbold{p}'^2}{2\MKc}+|\mbold{p}-\mbold{p}'|\right]^{-1}\nn\\
&&\times \frac{1}{4}\left[(\mbold{p}+\mbold{p}')^2-\frac{(\mbold{p}^2-{\mbold{p}'}^2)^2}{(\mbold{p}-\mbold{p}')^2}\right],
  \end{eqnarray}
is a inhomogeneous function in the parameter $\delta$, and to the accuracy we
are working required at leading order in $\delta$
only, 
\begin{equation}
    G_\gamma(\mbold{p},\mbold{p}') =\frac{1}{4}\frac{1}{|\mbold{p}-\mbold{p}'|^2}\left[(\mbold{p}+\mbold{p}')^2-\frac{(\mbold{p}^2-{\mbold{p}'}^2)^2}{(\mbold{p}-\mbold{p}')^2}\right]+\cdots
 \end{equation}
In order to evaluate the one-photon
exchange contributions, we replace the terms
$\sim\mbold{p}^2\Psi^*_{nl}(\mbold{p})$ and
$\sim\mbold{p'}^2\Psi_{nl}(\mbold{p'})$ by making use of the Schr\"odinger equation,
\begin{equation}
  \left[\mbold{p}^2+\frac{\alpha^2\mu_+^2}{n^2}\right]\Psi_{nl}(\mbold{p})=8\pi\alpha\mu_+\int \dnu(\mbold{q})\frac{1}{|\mbold{p}-\mbold{q}|^2}\Psi_{nl}(\mbold{q}).
\end{equation}
Further, we use the Fourier transform of $|\mbold{p}-\mbold{p}'|^{-2}$,
$|\mbold{p}-\mbold{q}|^{-2}$ and $|\mbold{p}'-\mbold{q}|^{-2}$ to express
the wave functions in coordinate space. The one-photon exchange contribution
now reads at order $\alpha^4$,
\begin{equation}
  \frac{\pi\alpha}{\mu_+\Sigma_+}|\Psi_{n0}(\mbold{x}=0)|^2+\frac{\alpha^3\mu_+}{n^2\Sigma_+}\langle
  r^{-1}\rangle-\frac{3\alpha^2}{2\Sigma_+}\langle
  r^{-2}\rangle,
\label{Egamma}
\end{equation}
where the expectation values are defined as
\begin{equation}
  \langle r^{-k}\rangle = \int
  d^3\mbold{r}\,\Psi^*_{nl}(\mbold{r})\frac{1}{|\mbold{r}|^k}\Psi_{nl}(\mbold{r}), 
\quad k=1,2.
\end{equation}
The electromagnetic energy shift at order $\alpha^4$ yields
\begin{eqnarray}
   \Delta E^{\rm em}_{nl}
   &=&\frac{\alpha^4\mu_+}{n^3}\left(1-\frac{3\mu_+}{\Sigma_+}\right)\left[\frac{3}{8n}-\frac{1}{2l+1}\right]+\frac{4\alpha^4\mu_+^3\lambda}{n^3}\delta_{l0}\nn\\
&& +\frac{\alpha^4\mu_+^2}{\Sigma_+}\left[\frac{1}{n^3}\delta_{l0}+\frac{1}{n^4}-\frac{3}{n^3(2l+1)}\right]+\Order{(\alpha^5{\rm
   ln}\alpha)}.
\label{DeltaEem}
\end{eqnarray}
Here, the first terms is generated by the mass insertions, the second contains
the finite size effects and the last stems from the one-photon exchange
contributions (\ref{Egamma}). For $n$ and $l$ arbitrary, we get the same
result for the pure QED contributions as
Ref. \cite{Nandy:rj,Todorov:1970gr,Austen:te}. 
Further the
formula for the ground-state agrees with the result obtained in Ref. \cite{Lyubovitskij:2000kk} for the electromagnetic energy
shift of the $\pi^-p$ atom, if we replace $\lambda$ through the corresponding
quantity in pionic hydrogen.
\begin{figure}[t]
\begin{center}
\leavevmode
\makebox{\includegraphics[height=2.3cm]{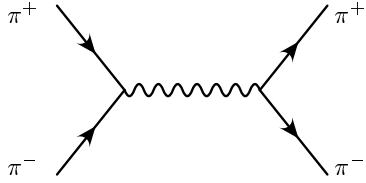}}
\caption{Annihilation diagram $\pi^+ \pi^-\rightarrow\pi^+ \pi^-$ at order
  $\alpha$ in the
relativistic theory.}
\label{fig: pipannihilation}
\end{center}
\end{figure}

To analyze the electromagnetic energy splittings of pionium, we need to
construct an effective Lagrangian that describes the relativistic
$\pi^+\pi^-\rightarrow \pi^+\pi^-$ amplitude at threshold correctly up to and
including order $\alpha$. The annihilation graph showed in
Figure \ref{fig: pipannihilation} corresponds to a local four pion interaction in the
non-relativistic Lagrangian. However, the corresponding relativistic scattering
matrix element vanishes at threshold. We may thus obtain the
electromagnetic energy-level shift from Eq. (\ref{DeltaEem}) by simply substituting $\mu_+\rightarrow \Mpic/2$,
$\Sigma_+\rightarrow 2\Mpic$ and $\lambda\rightarrow 1/3\langle r_{\pi^+}^2 \rangle$,
  \begin{equation}
    \Delta E^{\rm em}_{\pi, nl} =
    \alpha^4\Mpic\left[\frac{\delta_{l0}}{8n^3}+\frac{11}{64n^4}-\frac{1}{2n^3(2l+1)}\right]+\frac{\alpha^4\Mpic^3\langle r_{\pi^+}^2 \rangle}{6n^3}\delta_{l0}+\Order{(\alpha^5{\rm
   ln}\alpha)}.
\label{DeltaEempi}
\end{equation}
\subsection[Decay channels of the $\pi K$ atom]{Decay channels of the \boldmath{$\pi K$} atom}
\label{section: intStates}
Next we discuss the contributions from other decay channels to the decay width
of the $\pi^- K^+$ atom. As already mentioned in Section \ref{section:
  overview}, for S-states the only possible decay
channels are $K^0 + n_1\gamma+n_2e^+e^-$ and $\pi^0 K^0+n_1\gamma+n_2e^+e^-$,
where $n_1+n_2>0$. Here $n_1$ denotes the number of photons and $n_2$ the
number of $e^+e^-$ pairs. The decay widths into $\pi^- K^++n_1\gamma+n_2
e^+e^-$ vanish due to lack of phase space. Moreover, radiative transitions\footnote{Transitions between S-states with the
  simultaneous emission of two photons are not forbidden. However, they are
  suppressed by a factor of $\alpha^8$.}
with the emission of one photon are forbidden
between two states with $l=0$. For the 2P-state of the $\pi^-
K^+$ atom on the contrary, the main annihilation mechanism is the $2p-1s$ radiative transition into
the ground-state, followed by the decay into $\pi^0 K^0$. 

To investigate the  decays into $K^0+n\gamma$, $n=1,2$ we have to extend the
Lagrangian in Eqs. (\ref{Lagr}) and (\ref{Lagr2}) to include terms with odd intrinsic parity, such as
\begin{equation}
  \Lagr_{\rm A} = e D_1\mbold{B}\cdot(\pim^\dagger \DLR\Kp^\dagger\Kn+{\rm
  h.c})+e^2D_2\mbold{E}\cdot \mbold{B}(\pim^\dagger\Kp^\dagger\Kn+{\rm
  h.c})+\cdots,
\label{LagrA}
\end{equation}
where $u\DLR v \doteq u\mbold{D}v-v\mbold{D}u$. The covariant derivative
$\mbold{D}$ is specified in appendix \ref{appendix: Lagr}, $\mbold{E}$ denotes
the electric and $\mbold{B}$ the magnetic field. The couplings $D_1$ and $D_2$
are real and may be determined through matching with the chiral expansion of the
relativistic amplitudes. In the relativistic theory, the 
$\pi^+K^-K^02\gamma$ vertex is contained
in the Wess-Zumino-Witten term \cite{Wess:yu}. The $\pi^+K^-K^0\gamma$ interaction occurs not until the odd intrinsic parity sector of the
ChPT Lagrangian at $\Order{(p^6)}$ \cite{Bijnens:2001bb}. 
The such extended Hamiltonian is
hermitian and the operator $\mathbold{\bar{\tau}}$ obeys the unitarity condition,
\begin{equation}
  \mathbold{\bar{\tau}}(z)- \mathbold{\bar{\tau}}^\dagger(z) = -2\pi i
  \mathbold{\bar{\tau}}(z)\bar{\delta}(z-\mbold{H}_0-\mbold{H}_{\sss \rm C})\mathbold{\bar{\tau}}^\dagger(z).
\label{unitaritycond}
\end{equation} 
The symbol $\bar{\delta}$ is understood as follows: in order to evaluate the
right-hand side of the equation, we insert a complete set of
eigenstates
$(\mbold{H}_0+\mbold{H}_{\sss \rm C})$$\mid\!\beta\rangle=E_\beta\mid\!\beta\rangle$,
omitting the $n$-th Coulomb eigenstate of the $\pi^- K^+$ atom. This implies for the total decay width:
\begin{equation}
  \Gamma_{nl} = \sum_\beta \Gamma_{nl, \beta}, 
\end{equation}
where
\begin{eqnarray}
\Gamma_{nl, \beta} &=& \Int d p_\beta \, \dnu(\mbold{P})2\pi \delta(z-E_\beta)\langle\Psi_{nl},\mbold{P}\!\mid
\mathbold{\bar{\tau}}(z)\mid\!\beta\rangle\nn\\
&&\times\langle\beta\!\mid
\mathbold{\bar{\tau}}^\dagger(z)\mid\Psi_{nl},\mbold{0}\rangle,
\label{unitarity}
\end{eqnarray}
and $z$ is the solution of the master equation (\ref{masterformula}). Here,
$dp_\beta$ denotes the phase space integral over the intermediate state $|\beta\rangle$. At the
accuracy we are considering, we may use $z=E_n$. 
In the following, we estimate the order of the various decays using this
formula. As an illustration, we start with the decay into $\pi^0 K^0$.
The relative 3-momenta of the $\pi^-K^+$ pairs $\mbold{p}$ and $\mbold{p'}$ count
as order $\delta$ and we have
\begin{equation}
  \dnu(\mbold{p})\dnu(\mbold{p}')\Psi_{nl}^*(\mbold{p})\Psi_{nl}(\mbold{p}') =
  \Order{(\delta^3)}.
\label{wavefunction}
\end{equation}
As can be read off from the energy delta function, 
the outgoing $\pi^0$ and $K^0$ 3-momenta $\mbold{p}_3$ and $\mbold{p}_4$ count
as order $\delta^{1/2}$. This leads to a phase space suppression factor of
order $\delta^{1/2}$,
\begin{equation}
  \dnu(\mbold{p}_3)\dnu(\mbold{p}_4)\delta^3(\mbold{p}_3+\mbold{p}_4)\delta\left(E_n-E_{\pi^0K^0}\right)
  =\Order{(\delta^{1/2})},
\label{phasespace}
\end{equation}
where
\begin{equation}
  E_{\pi^0K^0} = \Sigma_0+\frac{\mbold{p}_3^2}{2\Mpin}+\frac{\mbold{p}_4^2}{2\MKn}.
\end{equation}
Further, the reduced matrix element $_+\!( \mbold{p}\!\mid
\mathbold{\bar{\tau}}(E_n)\mid\!\mbold{p}_3)_0$ is of $\Order{(1)}$
and the S-wave decay width thus starts at order $\delta^{7/2}$. The relation
(\ref{unitarity}) allows one to rather straightforwardly rederive the next-to-leading order formula for the decay width of the
S-states. In order not to interrupt the argument, we relegate the relevant
calculation to appendix \ref{appendix: unitarity condition}, and continue here
with the discussion of the radiative decay into $\pi^0 K^0+\gamma$.
\begin{figure}[ttbp]
\begin{center}
\leavevmode
\makebox{\includegraphics[height=7cm]{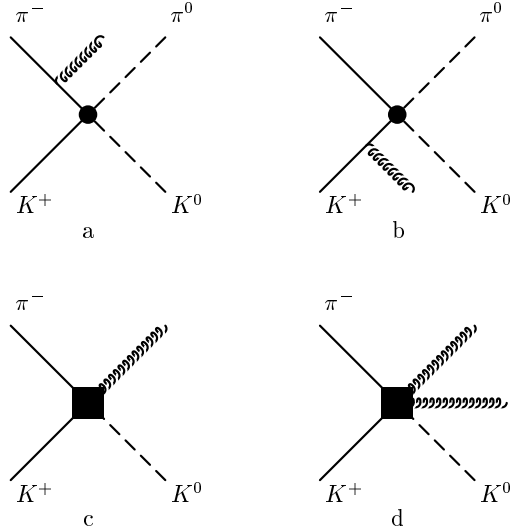}}
\caption{Leading order contributions to the decays into $\pi^0K^0+\gamma$ and
  $K^0 +n\gamma$, $n=1,2$. The dot denotes the coupling $C_2$, while the box stands for the  couplings
  $D_1$ and $D_2$ in $\Lagr_{\rm A}$. The twisted lines denote transverse photons.}
\label{fig: intStates}
\end{center}
\end{figure}
The outgoing $\pi^0$ and $K^0$ 3-momenta again count as
$\Order{(\delta^{1/2})}$, while the outgoing photon 3-momentum $\mbold{k}$ is of order
$\delta$. This can by seen by performing the phase space integrations over $\mbold{p}_3$, $\mbold{p}_4$ and $\mbold{k}$ explicitly. In total, the phase space suppression factor amounts to $\delta^{5/2}$,
\begin{equation}
  \dnu(\mbold{p}_3)\dnu(\mbold{p}_4)\frac{d^3\mbold{k}}{2\mid \!\mbold{k}\!\mid}\delta^3(\mbold{p}_3+\mbold{p}_4+\mbold{k})\delta\left(E_n-E_{\pi^0K^0\gamma}\right)=\Order{(\delta^{5/2})},
\end{equation}
with
\begin{equation}
  E_{\pi^0K^0\gamma}=\Sigma_0+\frac{\mbold{p}_3^2}{2\Mpin}+\frac{\mbold{p}_4^2}{2\MKn}+\mid \!\mbold{k}\!\mid.
\end{equation}
The leading order contribution stems from Figure \ref{fig: intStates}(a) and \ref{fig:
  intStates}(b). The corresponding reduced matrix element is given by
\begin{eqnarray}
  {}_+(\mbold{p}, -\mbold{p}\!\mid
  \mbold{V}\mathbold{\bar{G}}^n_{\rm \sss C}\mbold{V}\mid\!\mbold{p}_3,\mbold{p}_4, \mbold{k},
  \lambda)_0&=& -e C_2
  \mbold{p}\cdot\mathbold{\epsilon}(k,\lambda)
  \left[f(\Mpic,\mbold{p}^2,-\mbold{p}\cdot\mbold{k})\right.\nn\\
&&+\left. f(\MKc,\mbold{p}^2,\mbold{p}\cdot\mbold{k})\right]+\cdots,
\end{eqnarray}
where 
\begin{eqnarray}
f(M,\mbold{p}^2,\pm \mbold{p}\cdot\mbold{k}) &=&
  \frac{1}{M}\left[\frac{\alpha^2\mu_+}{2n^2}+\frac{\mbold{p}^2}{2\mu_+}+\frac{\mbold{k}^2}{2M}+\mid
  \!\mbold{k}\!\mid\pm \frac{\mbold{p}\cdot\mbold{k}}{M}\right]^{-1}.
\end{eqnarray}
This matrix element is of order $\delta^{1/2}$ which implies that the decay width $\Gamma_{\pi^0K^0\gamma}$
starts at order $\delta^{13/2}$. However,  this contribution 
  vanishes after performing the integrations over $\mbold{p}$ and $\mbold{p'}$.

Next, we consider the decay into $K^0+n\gamma$, $n=1,2$ (see Figure \ref{fig:
  intStates}(c) and \ref{fig:
  intStates}(d)). Here, the outgoing $K^0$ and photon 3-momenta belong to the
hard scale and thus count as $\Order{(1)}$. For $\pi^-K^+\rightarrow K^0+\gamma$,
  the Lagrangian (\ref{LagrA})
  leads to a reduced matrix element of order $\delta^{3/2}$,
\begin{equation}
  {}_+(\mbold{p}, -\mbold{p}\mid \mbold{V}\mid \mbold{p}_4, \mbold{k}, \lambda)_0 =
  2e D_1 \mbold{p}\cdot \left(\mbold{k}\times\mathbold{\epsilon}(k,\lambda)\right)+\cdots.
\label{matrixA}
\end{equation}
Naive power counting implies that the decay width into $K^0+\gamma$ starts at order
  $\delta^6$. The matrix element (\ref{matrixA}) is odd in $\mbold{p}$ and
the S-wave decay width therefore even more suppressed, while $\Gamma_{n1, K^0\gamma}$ starts at order $\delta^6$. 

For the transition $\pi^-K^+\rightarrow K^0 +2\gamma$, we get from the Lagrangian (\ref{LagrA}) a local matrix element of order $\delta$,
\begin{eqnarray}
   {}_+(\mbold{p}, -\mbold{p}\mid \mbold{V}\mid \mbold{p}_4, \mbold{k}_1,
   \lambda_1, \mbold{k}_2, \lambda_2)_0&=&
   e^2D_2k^0_1\mathbold{\epsilon}(k_1,\lambda_1)\cdot\left[\mbold{k}_2\times\mathbold{\epsilon}(k_2,\lambda_2)\right]\nn\\
   &&-2e^2D_1\mathbold{\epsilon}(k_1,\lambda_1)\cdot\left[\mbold{k}_2\times\mathbold{\epsilon}(k_2,\lambda_2)\right]\nn\\
&&+k_1\leftrightarrow k_2,\lambda_1\leftrightarrow\lambda_2,
\label{matrixA2}
  \end{eqnarray}
and the decay width of the S-states into $K^0+2\gamma$ thus starts at order
$\delta^5$. For the P-wave decay width into $K^0+2\gamma$ this contributions
vanishes, because the matrix element in Eq. (\ref{matrixA2}) is $\mbold{p}$ independent.  

Processes with a higher number of photons may be treated in a analogous
manner. We expect them -
using power counting arguments - to be even more suppressed.
Since hard processes such as $K^0 +n_1\gamma+n_2e^+e^-$, $n_1+n_2>0$ do not contribute to the decay
width at order
$\delta^{9/2}$, we may assume that all couplings in the non-relativistic
Lagrangian in Eqs. (\ref{freeLagr}), (\ref{Lagr}) and (\ref{Lagr2}) are
real. The total S-wave decay width of the $\pi^- K^+$ atom amounts to  
\begin{equation}
  \Gamma_{n0} = \Gamma_{n0, \pi_0 K_0}+ \Order{(\delta^5)}.
\end{equation}
The $\pi^- K^+$ atom in the 2P-state on the other hand decays predominantly
through the radiative transition into the ground-state. To evaluate this
transition, we insert the ground-state plus one photon into
Eq. (\ref{unitarity}). 
Here the photon 3-momentum $\mbold{k}$ counts as order $\delta^2$, as can be
read off from the energy delta function $\delta(E_2-E_1-|\mbold{k}|)$. At
leading order, we get
for the spontaneous $2p-1s$ transition the well-known expression, see e.g. 
Ref. \cite{Bethe}
\begin{equation}
  \Gamma_{21} = \left(\tfrac{2}{3}\right)^8\alpha^5\mu_++\cdots
\label{transition}
\end{equation}
The result is of order $\alpha^5$ and given numerically in Table \ref{table:
  numericsK}. The first subleading decay mode of the 2P-state starts at order $\delta^6$ with the odd intrinsic parity decay into $K^0+\gamma$. The P-wave decay width
into $\pi^0 K^0$ in Eq. (\ref{GammaP}) is of
order $\delta^{13/2}$ and suppressed with respect to radiative $2p-1s$ transition by a factor of
$10^{-7}$.
\setcounter{equation}{0}
\section{Vacuum polarization}
\label{section: vacpol}
What remains to be added are the contributions coming from the
electron vacuum polarization. The calculation of these corrections within a
non-relativistic Lagrangian approach has been performed in
Ref. \cite{Eiras:2000rh}. In our framework, the contributions due to vacuum
polarization arise formally at higher
order in $\alpha$. However, they are amplified by powers of
the coefficient $\mu_+/m_e$, where $m_e$ denotes the electron mass. 
To the accuracy considered here, the only effect of the vacuum polarization
of the electron is a modification of the Coulomb potential $\mbold{H}_{\sss \rm C}\rightarrow\mbold{H}_{\sss \rm
  C}+\mbold{H}_{\rm vac}$, with
\begin{equation}
  {}_+(\mbold{p}\!\mid
  \mbold{H}_{\rm vac}\mid\!\mbold{q})_+ =
  -\frac{4\alpha^2}{3}\int_{4m_e^2}^\infty
  \frac{ds}{s+(\mbold{p}-\mbold{q})^2}\frac{1}{s}\left[1+\frac{2m_e^2}{s}\right]\left[1- \frac{4m_e^2}{s}\right]^{1/2}.
\end{equation}
The vacuum polarization leads to an electromagnetic energy shift
evaluated in Refs. \cite{Eiras:2000rh,Lyubovitskij:2000kk,vacpol},
\begin{equation}
  \Delta E^{\rm vac}_{nl} = \left(\Psi_{nl}\!\mid\mbold{H}_{\rm
  vac}\mid\!\Psi_{nl}\right).
\label{DeltaEvac}
\end{equation}
For the first two energy-level shifts of pionium\footnote{For pionium, the
  electromagnetic energy shift due to vacuum polarization is denoted by $\Delta E^{\rm
  vac}_{\pi, nl}$.} and the $\pi^\pm K^\mp$ atom, $\Delta E^{\rm
  vac}_{\pi, nl}$ and $\Delta E^{\rm
  vac}_{nl}$ are
  given numerically in Table \ref{table: numericspi} and
\ref{table: numericsK}.
Formally of order $\alpha^{2l+5}$, this contribution is numerically sizeable
due to its large coefficient containing $(\mu_+/m_e)^{2l+2}$. 

The vacuum polarization also interferes with strong interactions and
contributes to the decay width and to the strong energy shift. This
can be seen by inserting the modified Coulomb potential into the master equation (\ref{masterformula}). For the spectrum and
the width of the $\pi^\pm K^\mp$ atom, we get
\begin{eqnarray}
  \Gamma_{n0} &=&
  \frac{8\alpha^3\mu_+^2}{n^3}p_n^*(a_0^-)^2\left(1+\delta_{{\sss K},
  n}+\delta^{\rm vac}_{{\sss K},
  n}\right),\nn\\
  \Delta E_{n0}^{\rm h} &=&
  -\frac{2\alpha^3\mu_+^2}{n^3}(a_0^++a_0^-)\left(1+\delta_{{\sss K},
  n}'+\delta^{\rm vac}_{{\sss K},
  n}\right).
\label{Kvac}
\end{eqnarray}
What concerns pionium, the decay width and strong energy shift are modified, according to 
\begin{eqnarray}
 \Gamma_{\pi, n0} &=& \frac{2\alpha^3}{9n^3}p_{\pi,
 n}^*(a_0^0-a_0^2)^2\left(1+\delta_{\pi, n}+\delta^{\rm vac}_{\pi, n}\right),\nn\\
\Delta E_{\pi, n0}^{\rm h}
  &=&-\frac{\alpha^3\Mpic}{6n^3}(2a_0^0+a_0^2)\left(1+\delta_{\pi,
  n}'+\delta^{\rm vac}_{\pi, n}\right).
\label{pivac}
\end{eqnarray}
The correction $\delta^{\rm vac}_{h,n}$, $h = \pi, K$ is proportional to the
change in the Coulomb wave function \cite{Eiras:2000rh} of the bound system due to vacuum polarization,
\begin{equation}
  \delta^{\rm vac}_{h,
  n}=\frac{2\delta \Psi_{n0}(\mbold{x}=0)}{\Psi_{n0}(\mbold{x}=0)}.
\label{deltavac}
\end{equation}
Here, $\Psi_{nl}$ stands for a generic Coulomb wave function and $h=\pi, K$. For the
  ground-state, this result is contained in Table II of
  Ref. \cite{Eiras:2000rh}. Formally, $\delta^{\rm vac}_{h,n}$ is of order $\alpha^2$, but
  enhanced because of the large coefficient containing $\mu_+/m_e$.
\setcounter{equation}{0}
\section{Numerics}
\label{section: numerics}
\begin{table}[t]
\begin{center}
\begin{tabular}{|l|r|r|r|r|}\hline
\rule{0mm}{3.5mm}
& $10^2\delta_{h, 1}$ & $10^2\delta_{h, 2}$ & $10^2\delta_{h, 1}'$ & $10^2\delta_{h, 2}'$\\\hline
$\pi^+\pi^-$ atom &$5.8\pm 1.2$&$5.6\pm1.2$ &$6.2\pm 1.2$ &$6.1\pm 1.2$ \\
$\pi^\pm K^\mp$ atom &$4.0\pm2.2$&$3.8\pm2.2$ &$1.7\pm2.2$ &$1.5\pm2.2$ \\\hline
\end{tabular}
\end{center}
 \medskip
 \caption{Next-to-leading order corrections to the Deser-type formulae. \label{table: delta}}
\end{table}
In the numerical evaluation of the widths and energy shifts of the
$\pi^+\pi^-$ and $\pi^\pm K^\mp$
atoms, we use the following numbers: 
The $\pi\pi$ scattering lengths yield $a_0^0=
0.220\pm0.005$, $a_0^2 = -0.0444\pm0.0010$ and $a_1^1=(0.379\pm0.005)\cdot10^{-1}\Mpic^{-2}$
\cite{Colangelo:2000jc,Colangelo:2001df}. The correlation matrix for $a_0^0$ and $a_0^2$ is given in Ref. \cite{Colangelo:2001df}. For the isospin symmetry breaking corrections to the $\pi\pi$ threshold
amplitudes (\ref{Gammapi}) and (\ref{DeltaEpi}) at order $e^2p^2$, we use $\epsilon_\pi=(0.61\pm0.16)\cdot 10^{-2}$ and $\epsilon_\pi'=
(0.37\pm0.08)\cdot 10^{-2}$ \cite{Gasser:2001un,Knecht:2002gz}. The values for the $\pi K$ scattering
lengths are taken from the recent analysis of data and Roy-Steiner equations
\cite{Buettiker:2003pp}. The S-wave scattering lengths yield  $a_0^+=(0.045\pm
0.012)\Mpic^{-1}$, $a_0^-=(0.090\pm 0.005)\Mpic^{-1}$\cite{Buettiker:2003pp},
and for the P-waves
we use $a_1^{1/2}=(0.19\pm0.01)\cdot10^{-1}\Mpic^{-3}$ and $a_1^{3/2} = (0.65\pm0.44)\cdot10^{-3}\Mpic^{-3}$\cite{Buettiker:2003pp}. 
The correlation parameter for $a_0^+$ and $a_0^-$ is also given in
Ref. \cite{Buettiker:2003pp}.
The isospin breaking corrections to the $\pi K$ threshold amplitudes (\ref{eq:
  A}) and (\ref{eq: Ap}) have been
worked out in  Refs. \cite{Nehme:2001wa,Kubis:2001ij,Kubis} at $\Order(p^4,e^2p^2)$. Whereas the analytic expressions for
$\epsilon$ and $\epsilon'$ obtained in Refs. \cite{Nehme:2001wa,Kubis:2001ij,Kubis} are not identical,
the numerical values agree within the uncertainties quoted in \cite{Kubis}. In the following, we use \cite{Kubis} $\epsilon=(0.1\pm0.1)\cdot
10^{-2}M_{\pi^+}^{-1}$ and
  $\epsilon'=(0.1\pm0.3)\cdot10^{-2}M_{\pi^+}^{-1}$. For the charge radii of
  the pion and kaon, we take $\langle r^2_{\pi^+}\rangle =
(0.452\pm0.013)\,{\rm fm^2}$ and $\langle r^2_{\sss K^+}\rangle
=(0.363\pm0.072)\,{\rm fm^2}$ \cite{Bijnens:2002hp}. In the evaluation of the
uncertainties, we take into account the correlation between the $\pi\pi$, ($\pi K$) scattering lengths.

The isospin breaking corrections $\delta_{h,n}$ and $\delta_{h,n}'$, $h=\pi,
K$ to the widths (\ref{Gammanum}), (\ref{Gammanumpi}) and strong energy shifts
(\ref{DeltaEnum}), (\ref{DeltaEnumpi}) are listed in Table
\ref{table: delta}. The energy shift corrections $\delta_{{\sss K},n}'$ are smaller than in the case of pionium. This
discrepancy comes from the different size of the isospin breaking
contributions to the elastic one-particle irreducible $\pi\pi$ and $\pi K$ amplitudes. At leading order
in the chiral expansion, the isospin breaking part of the $\pi^-
K^+\rightarrow\pi^- K^+$ amplitude at threshold is suppressed by a factor of
$\Mpic/\MKc$ with respect to the corresponding quantity in $\pi\pi$
scattering. 

As described in
Section \ref{section: vacpol}, Eqs. (\ref{Kvac}) and (\ref{pivac}), the width and
strong energy shift are modified due to vacuum polarization, according to
\begin{equation}
  \delta_{h, n} \rightarrow \delta_{h, n}+ \delta_{h, n}^{\rm vac}, \quad \delta_{h, n}' \rightarrow \delta_{h, n}'+ \delta_{h, n}^{\rm vac},
\end{equation}
where $h=\pi, K$ and $\delta_{h, n}^{\rm vac}$ is defined in Eq. (\ref{deltavac}). 
For the ground-state, the corrections due to vacuum polarization yield $\delta^{\rm vac}_{{\sss
    K}, 1}=0.45\cdot 10^{-2}$ and $\delta^{\rm vac}_{\pi, 1}=0.31\cdot
10^{-2}$\cite{Eiras:2000rh}. However, for the numerical analysis of the width and the strong energy
shift, we may neglect the contributions from $\delta^{\rm vac}_{h, n}$, because
the uncertainties in $\delta_{h, n}$ and $\delta_{h, n}'$ are larger than
$\delta^{\rm vac}_{h, n}$ itself. 
\begin{table}[t]
\begin{center}
\begin{tabular}{|l|r|r|r|r|r}\hline
\rule{0mm}{3.5mm}
$\pi^+\pi^-$ atom & $\Delta E^{\rm em}_{\pi, nl}$[eV]&$\Delta E^{\rm vac}_{\pi,
  nl}$[eV]&$\Delta E^{\rm h}_{\pi, nl}$[eV]&$10^{15}\tau_{\pi, nl}$[s]\\\hline\rule{0mm}{3.5mm}
$n$=1, $l$=0&$-0.065$&$-0.942$&$-3.8\pm 0.1$ &$2.9\pm0.1$\\
$n$=2, $l$=0&$-0.012$&$-0.111$&$-0.47\pm 0.01$&$23.3\pm0.7$\\
$n$=2, $l$=1&$-0.004$&$-0.004$&$\simeq-1\cdot10^{-6}$&$\simeq1.2\cdot10^{4}$\\\hline
\end{tabular}
\end{center}
 \medskip
 \caption{Numerical values for the energy shift and the lifetime of the
   $\pi^+\pi^-$ atom.\label{table: numericspi}}
\end{table}
\begin{table}[t]
\begin{center}
\begin{tabular}{|l|r|r|r|r|r|}\hline
\rule{0mm}{3.5mm}
$\pi^\pm K^\mp$ atom&$\Delta E^{\rm em}_{nl}$[eV]&$\Delta E^{\rm vac}_{nl}$[eV]&
$\Delta E^{\rm h}_{nl}$[eV]&
$10^{15}\tau_{nl}$[s]\\\hline\rule{0mm}{3.5mm}
$n$=1, $l$=0&$-0.095$&$-2.56$&$-9.0\pm1.1$&$3.7\pm0.4$\\
$n$=2, $l$=0&$-0.019$&$-0.29$&$-1.1\pm0.1$&$29.4\pm3.3$\\
$n$=2, $l$=1&$-0.006$&$-0.02$&$\simeq -3\cdot10^{-6}$&$\simeq0.7\cdot 10^{4}$\\\hline
\end{tabular}
\end{center}
 \medskip
 \caption{Numerical values for the energy shift and the lifetime of the $\pi^\pm K^\mp$ atom.\label{table: numericsK}}
\end{table}

For the first two states of the $\pi^+\pi^-$ and $\pi^\pm K^\mp$ atoms, the numerical values for the lifetime $\tau_{nl} \doteq \Gamma^{-1}_{nl}$,
($\tau_{\pi, nl} \doteq \Gamma^{-1}_{\pi, nl}$) and the energy shifts are listed in Table \ref{table:
  numericspi} and \ref{table:
  numericsK}. The numbers for the lifetime and strong energy shifts of the
S-states are valid at next-to-leading order in isospin
symmetry breaking. The bulk part in
the uncertainties of these quantities is due to the
uncertainties in the corresponding scattering lengths. For the lifetime of the
2P-state, the numerical values are valid at leading order only, and determined
by the $2p-1s$ radiative transition in Eq. (\ref{transition}) \cite{Nemenov:wz}.   
 
The energy-level shift due to vacuum polarization $\Delta
E^{\rm vac}_{nl}$, ($\Delta
E^{\rm vac}_{\pi,nl}$) \cite{Eiras:2000rh,vacpol} is specified in
Eq. (\ref{DeltaEvac}). Formally of higher order in $\alpha$, this contribution
is numerically sizable. We do not display the error bars for the
electromagnetic energy shifts. At order $\alpha^4$, they come from the uncertainties in the
charge radii of the pion and kaon only. In the case of pionium, the uncertainties of $\Delta
E_{\pi, 10}^{\rm em}$ at order $\alpha^4$ amount to about $0.7\%$, while for
the $\pi^\pm K^\mp$ atom  $\Delta E_{10}^{\rm em}$ is known at the $5\%$ level. To estimate the order of magnitude of the electromagnetic corrections at higher
order, we may compare with positronium. Here, the $\alpha^5$ and $\alpha^5
\,{\rm ln}\alpha$ corrections \cite{Karplus:1952wp} amount to about $2\%$ with
respect to the $\alpha^4$ contributions. 
 
Both, the electromagnetic and vacuum polarization contributions to the energy
shift are known to a high accuracy. A future precision measurement of the
energy splitting between the $n$S and $n$P states \cite{Nemenov:vp} will
therefore allow one to extract the
strong S-wave energy shift in Eq. (\ref{DeltaEnum}), and to determine the
combination $a_0^+ + a_0^-$ of the $\pi K$ scattering lengths. The $2s-2p$
energy splitting yields   
\begin{eqnarray}
  \Delta E_{2s-2p} &=& \Delta E_{20}^{\rm h}+\Delta E_{20}^{\rm em}-\Delta
  E_{21}^{\rm em}+\Delta E_{20}^{\rm vac}-\Delta E_{21}^{\rm vac}\nn\\
&=&-1.4 \pm 0.1\,{\rm eV}.
\label{DeltaE2}
\end{eqnarray}
The uncertainty displayed is the one in $\Delta E_{2}^{\rm h}$ only. To the
accuracy we are working, we may neglect the strong shift in the $2$P state, it
is suppressed by the power of $\alpha^5$. For pionium the energy splitting
between the 2S and 2P states reads
\begin{eqnarray}
  \Delta E_{\pi, 2s-2p} &=& \Delta E_{\pi, 20}^{\rm h}+\Delta E_{\pi, 20}^{\rm
  em}-\Delta E_{\pi, 21}^{\rm em}+\Delta E_{\pi, 20}^{\rm vac}-\Delta E_{\pi,
  21}^{\rm vac}\nn\\
&=&-0.59\pm 0.01\,{\rm eV}.
\label{DeltaE2pi}
\end{eqnarray}
Again the uncertainty displayed is the one in $\Delta E_{\pi, 2}^{\rm h}$ only
and we may neglect contributions from the strong shift in the $2$P state.

As an illustration, we alternatively use the ChPT predictions for the $\pi
K$ scattering lengths \cite{Bernard:1990kx,Nehme:2001wa} in the
numerical evaluation of the lifetime.
The ChPT predictions yield at order $p^4$,
$a_0^+=(0.032\pm0.016)\Mpic^{-1}$ and $a_0^-=(0.079\pm0.001)\Mpic^{-1}$
\cite{Buettiker:2003pp}. Here, the errors include the
uncertainties in the values of the input parameters only. The uncertainty in
$a_0^-$ is remarkably small, because the isospin odd scattering length involves at
$\Order{(p^4)}$ a single low--energy constant $L_5^r$ \cite{Nehme:2001wa}. On
the other hand, $a_0^+$ contains apart from the combination $2L_6^r+L_8^r$
five further coupling constants \cite{Nehme:2001wa}, which are enhanced by one
power of $\MKc/\Mpic$ with respect to the counterterm in $a_0^-$. 
For $a_0^-$, the two-loop
correction has to be rather substantial, in order to reproduce the central
value of the Roy-Steiner evaluation. Very recently, the chiral expansion of
the $\pi K$ scattering
amplitude at next-to-next-to-leading order became available \cite{Bijnens:2004bu}. According to
the preliminary numerical study performed in Ref. \cite{Bijnens:2004bu},
the S-wave
scattering lengths are at order $p^6$ in reasonable agreement with the
Roy-Steiner evaluation \cite{Buettiker:2003pp}. The $\Order(p^4)$ ChPT prediction for the lifetime of the $\pi^\pm K^\mp$ atom in its
ground-state is
\begin{equation}
  \tau_{10} = 4.7\cdot 10^{-15}{\rm s}, \quad  \textrm{ChPT}, [\Order(p^4)],
\end{equation}
whereas
  \begin{equation}
    \tau_{10} = (3.7 \pm 0.4)\cdot 10^{-15}{\rm s}, \quad  \textrm{Roy-Steiner}.
  \end{equation}
The ChPT prediction is valid at next-to-leading order in isospin symmetry
breaking and up to and including $\Order{(p^4)}$ in the chiral expansion. 
\setcounter{equation}{0}
\section{Summary and Outlook}
We have considered the spectra and decays of $\pi^+\pi^-$ and $\pi^\pm K^\mp$
atoms in the framework of QCD $+$ QED. We
evaluated the corrections to the Deser-type formulae for the width and the
energy shift - valid at next-to-leading order in isospin symmetry breaking -
within a non-relativistic effective field theory. It is convenient to
introduce a different book keeping for the $\pi^+\pi^-$ and $\pi^\pm K^\mp$
atoms. What concerns pionium, we count $\alpha$ and $(m_u-m_d)^2$ as small quantities of order $\delta$,
in the case of the $\pi^\pm K^\mp$ atom both $\alpha$ and $m_u-m_d$ are of order
$\delta$. The different counting schemes are due to the fact that in QCD, the
pion mass difference starts at $(m_u-m_d)^2$, while the kaon mass difference is
linear in $m_u-m_d$. 

Consider first the energy shifts that are split into an
electromagnetic and a strong part, according to Eq. (\ref{DeltaEtot}). 
The electromagnetic part in Eqs. (\ref{DeltaEem}) and
(\ref{DeltaEempi}) contains both pure QED contributions as well as finite
size effects due to the charge radii of the pion and kaon. 
The strong energy shift of the $\pi^- K^+$ atom is
proportional to the one-particle irreducible $\pi^- K^+\rightarrow\pi^- K^+$
scattering amplitude at threshold. In the isospin symmetry limit, the elastic threshold amplitude reduces to the sum of isospin even and odd
scattering lengths $a_0^++a_0^-$. The isospin breaking
contributions to the amplitude have been evaluated at
$\Order{(e^2p^2,p^4)}$\cite{Nehme:2001wa,Kubis} in the framework of ChPT. The result in
  Eq. (\ref{DeltaEnum}) displays the S-wave energy shift in terms of the sum $a_0^++a_0^-$, and a correction of order $\alpha$ and
  $m_u-m_d$. For the first two energy-level shifts, the isospin symmetry
  breaking correction modifies the leading order term at the $2\%$ level. The
  isospin even scattering length $a_0^+$ is sensitive to the combination of
  low--energy constants $2L_6^r+L_8^r$. The consequences of this observation
  for the SU(3)$\times$SU(3) quark condensate \cite{Descotes-Genon:1999uh}
  remain to be worked out. In the case of pionium, the strong energy shift displayed in
  Eq. (\ref{DeltaEnumpi}) is related to the $\pi\pi$ scattering lengths combination
  $2a_0^0+a_0^2$ and a correction of order $\alpha$ and $(m_u-m_d)^2$. For the
  first two energy-levels, these isospin symmetry breaking contributions
  amount to about $6\%$. 

 A future measurement of the energy splitting between the $2$S and $2$P state
 in the $\pi^+ \pi^-$, ($\pi^\pm K^\mp$) atom will allow one to extract the
 strong energy shift and to determine the
 scattering lengths combination $2a_0^0+a_0^2$, ($a_0^++a_0^-$). 
This is due to the fact that the 
 electromagnetic energy shifts are known to high accuracy and the strong P-wave
 energy shifts in Eqs. (\ref{DeltaEP}) and
 (\ref{DeltaEPpi}) are suppressed by
 the power of $\alpha^{5}$. However, another higher order correction - 
 generated by the vacuum
 polarization - is numerically sizable and contributes to the energy splitting
 in Eqs. (\ref{DeltaE2}) and (\ref{DeltaE2pi}). Formally of order $\alpha^{2l+5}$,
 this correction is enhanced due to its large coefficient containing
 $(\mu_+/m_e)^{2l+2}$.

We now turn to the decay widths of the $\pi^+\pi^-$ and $\pi^- K^+$ atoms. At leading and next-to leading order the $\pi^+\pi^-$ and $\pi^- K^+$ atoms
decay into $2\pi^0$ and $\pi^0 K^0$ exclusively. Aside from a
kinematical factor - the relativistic outgoing 3-momentum of the bound system
- their decay width can be expanded in powers of $\alpha$ and
$m_u-m_d$. By invoking ChPT, the result for the S-wave decay width
 of the $\pi^- K^+$ atom may be expressed in terms of the isospin odd scattering length $a_0^-$,
and an isospin symmetry breaking correction of order $\alpha$ and $m_u-m_d$,
see Eq. (\ref{Gammanum}). For the ground-state decay width, this correction
modifies the leading order Deser-type relation at the $4\%$ level. The
next-to-leading order result
for the S-wave decay width of pionium is given in Eq. (\ref{Gammanumpi}). The
expression for
the ground-state width agrees with the result obtained in
Refs. \cite{Jallouli:1997ux,Gasser:2001un,Gasser:1999vf}. 

For the 2P state of the $\pi^- K^+$, ($\pi^+\pi^-$) atom, the decay width
starts at order $\alpha^5$ with the $2p-1s$ radiative transition into
the ground-state, see Eq. (\ref{transition}). The P-wave decay width of the $\pi^- K^+$ atom into $\pi^0
K^0$ in Eq. (\ref{GammaP}) is suppressed by the power of $\delta^{13/2}$. For
pionium, the P-wave decay width into a pair of two neutral pions vanishes due to $C$ invariance.

We find it very exciting that in view to the beautiful work performed by our
experimental colleagues, we may expect that many of the above predictions
will be confronted with experimental data in a not too distant future. This
will certainly improve our knowledge of the low--energy structure of QCD. 
\section*{Acknowledgements}
I am grateful to J. Gasser for his help and advice throughout this work and
for reading the manuscript carefully.  
I thank P. B\"uttiker, R. Kaiser, B. Kubis, A. Rusetsky, J. Schacher and H. Sazdjian for useful discussions.
This work was supported in part by the Swiss National Science Foundation and by
 RTN, BBW-Contract N0.~01.0357 and EC-Contract HPRN--CT2002--00311 (EURIDICE).
\begin{appendix}
\renewcommand{\thesection}{\Alph{section}}
\renewcommand{\theequation}{\Alph{section}.\arabic{equation}}
\setcounter{equation}{0}
\section{Non-relativistic Lagrangian}
\label{appendix: Lagr}
The non-relativistic Lagrangian must be invariant under space rotation, $P$, $T$
and gauge transformations. We do not include terms corresponding to
transitions between sectors with different numbers of heavy fields (pions and kaons). These
interactions describe processes with an energy release at the hard scale. In
general, such decay processes are accounted for by
introducing complex couplings in the non-relativistic Lagrangian. However, as shown in
Section \ref{section: intStates}, intermediate states do not contribute to the decay
width at order $\delta^{9/2}$ and we may therefore assume the low--energy
couplings to be real. 

In the sector with one or two mesons, the non-relativistic effective
Lagrangian is given by
\begin{equation}
  \Lagr_{\sss \rm NR} = \Lagr_1+\Lagr_2^{(0)}+\Lagr_2^{(2)}+\cdots
\end{equation}
The first term contains the one-pion and one-kaon sectors
\begin{eqnarray}
  \Lagr_1 &=& \frac{1}{2}(\mbold{E}^2-\mbold{B}^2)+\pin^\dagger \Big( i \d_t
  -\Mpin+\frac{\Delta}{2\Mpin}+\frac{\Delta^2}{8\Mpin^3}+\cdots\Big)\pin\nn\\
&&+\Kn^\dagger \Big( i \d_t
  -\MKn+\frac{\Delta}{2\MKn}+\frac{\Delta^2}{8\MKn^3}+\cdots\Big)\Kn\nn\\
&&+ \sum_{\pm}\pi_\pm^\dagger \Big( i D_t
  -\Mpic+\frac{\mbold{D}^2}{2\Mpic}+\frac{\mbold{D}^4}{8\Mpic^3}+\cdots\Big)\pi_\pm\nn\\
&&+ \sum_{\pm} K_\pm^\dagger\Big( i D_t
  -\MKc+\frac{\mbold{D}^2}{2\MKc}+\frac{\mbold{D}^4}{8\MKc^3}+\cdots\Big)K_\pm\nn\\
&&\mp e\sum_{\pm}
  \mbold{D}(\mbold{E})\left(\frac{c_\pi}{6\Mpic^2}\pi_\pm^\dagger
  \pi_\pm+\frac{c_{\sss K}}{6\MKc^2}K_\pm^\dagger K_\pm+\cdots\right),
\label{freeLagr}
\end{eqnarray}
with electromagnetic charge $e$, $\mbold{E} = -\nabla A_0-\dot{\mbold{A}}$ and
$\mbold{B}
= \nabla\times \mbold{A}$. The covariant derivatives of the charged meson
fields are given by
\begin{eqnarray}
  && D_t \pi_\pm = \d_t \pi_\pm \mp i e A_0 \pi_\pm, \quad \mbold{D}\pi_\pm = \nabla
  \pi_\pm \pm i e \mbold{A}\pi_\pm,\nn\\
 && D_t K_\pm = \d_t K_\pm \mp i e A_0 K_\pm, \quad \mbold{D}K_\pm = \nabla K_\pm \pm i e \mbold{A}K_\pm.
\end{eqnarray}
The one-pion-one-kaon sector of total zero charge reads at lowest order
\begin{eqnarray}
  \Lagr_2^{(0)}&=& C_1\pim^\dagger\Kp^\dagger\pim\Kp +C_2\left(
  \pim^\dagger\Kp^\dagger\pin\Kn+\textrm{h.c}\right)+
  C_3\pin^\dagger\Kn^\dagger\pin\Kn.
\label{Lagr}
\end{eqnarray}
To evaluate the decay width and energy shifts of the $\pi^- K^+$ atom, we
need in addition the following terms with two covariant space derivatives\footnote{In the c.m. system and by the use of the equations of motion, we identify
  \begin{equation}
  \pim^\dagger \Kp^\dagger \pin \dLR \Kn = -4\mu_0(\Sigma_+-\Sigma_0)\pim^\dagger\Kp^\dagger\pin\Kn+\frac{\mu_0}{\mu_+}\pim^\dagger \DLR^2\Kp^\dagger\pin\Kn\nn.
  \end{equation}
} 
\begin{eqnarray}
   \Lagr_2^{(2)} &=&  C_4\left(\pim^\dagger\DLR^2\Kp^\dagger\pin\Kn+\textrm{h.c}\right)+C_5
  \left(\pim^\dagger\DLR^2\Kp^\dagger\pim\Kp+{\rm h.c}\right)\nn\\
&&+C_6(\pim^\dagger\pim)\mbold{D}^2(\Kp^\dagger\Kp)+
  C_7\left(\nabla\pim^\dagger\Kp^\dagger\nabla\pin\Kn+\textrm{h.c}\right)+\cdots,
\label{Lagr2}
\end{eqnarray}
where $u\DLR^2 v \doteq u\mbold{D}^2 v+v\mbold{D}^2 u$. We work in the center of
mass  system and thus omit terms proportional to the c.m. momentum. We do not display time derivatives, for on-shell
matrices they can be eliminated by the use of the equations of motion. The
parameters $M_{\pi^i}$ ($M_{K^i}$) denote the physical pion, (kaon) masses --
there is no mass renormalization in the non-relativistic theory, see Section
\ref{section: matching}. 

We work in the Coulomb gauge and eliminate the $A^0$ component of the photon
field by the use of the equations of motion. At the accuracy we are
considering, the term linear in $\mbold{D}(\mbold{E})$ in Eq. (\ref{freeLagr})
then reduces to a local interaction which modifies the low--energy coupling $C_1$,  
\begin{eqnarray}
  C_1' &=& C_1- e^2\lambda,\nn\\
\lambda &=& \frac{c_\pi}{6\Mpic^2}+\frac{c_{\sss K}}{6\MKc^2}.
\end{eqnarray}
It is sufficient to match the non-relativistic couplings $c_\pi$ and $c_{\sss
  K}$ at order $\delta^0$. We therefore consider the pion and kaon
  electromagnetic form-factors in the external field $A_\mu$. The results of
  the matching yield
\begin{equation}
  c_\pi = \Mpic^2 \langle r^2_\pi \rangle, \quad  c_{\sss K} = \MKc^2 \langle
  r^2_{\sss K} \rangle,
\label{radius}
\end{equation}
where $r_\pi$ and $r_{\sss K}$ denote the charge radii of the charged pion
and kaon, respectively. The remaining low energy constants $C_i$, $i=1,\dots
7$ may be determined
through matching the $\pi K$ amplitude in the vicinity of the threshold for different channels,
see Section \ref{section: matching}. 
\setcounter{equation}{0}
\section{Relativistic scattering amplitudes}
\label{appendix: relAmplitude} 
First, we consider the $S = 1$ processes
 \begin{equation}
   \pi^- K^+ \rightarrow \pi^- K^+, \quad \pi^- K^+ \rightarrow \pi^0 K^0,
   \quad \pi^0 K^0 \rightarrow \pi^0 K^0,
 \end{equation}
in the isospin symmetry limit. The decomposition into amplitudes with definite isospin yields
 \begin{eqnarray}
   T^{\pm;\pm}&=& \frac{1}{3}\left[T^{\scriptscriptstyle
       3/2}(s,t)+2T^{\scriptscriptstyle 1/2}(s,t)\right],\nn\\
   T^{00;\pm} &=& \frac{\sqrt{2}}{3}\left[ T^{\scriptscriptstyle
   3/2}(s,t)-T^{\scriptscriptstyle 1/2}(s,t)\right],\nn\\
T^{00;00}&=&  \frac{1}{3}\left[2T^{\scriptscriptstyle
       3/2}(s,t)+T^{\scriptscriptstyle 1/2}(s,t)\right].
 \end{eqnarray}
  The isospin $I = 1/2$ and $I=3/2$ components are related via
  \begin{equation}
    T^{\scriptscriptstyle 1/2}(s,t,u) = \frac{3}{2}T^{\scriptscriptstyle 3/2}(u,t,s)-\frac{1}{2}T^{\scriptscriptstyle 3/2}(s,t,u).
  \end{equation}
 The $T^+$, ($T^-$) amplitude
 \begin{eqnarray}
   T^+(s,t) &=& \frac{1}{3}\left[T^{\scriptscriptstyle
       1/2}(s,t)+2T^{\scriptscriptstyle 3/2}(s,t)\right],\nn\\
   T^-(s,t) &=& \frac{1}{3}\left[T^{\scriptscriptstyle
       1/2}(s,t)-T^{\scriptscriptstyle 3/2}(s,t)\right],
 \end{eqnarray}
 is even (odd) under $s \leftrightarrow u$ crossing.
In the s-channel, the decomposition into partial waves reads
 \begin{equation}
   T^I(s,t) =
   16\pi\sum_{l=0}^\infty(2l+1)t^I_l(s)P_l({\rm cos}\,\theta),
 \end{equation}
 where $s = [\w{{\pi^+}}{\mbold{p}}+\w{{K^+}}{\mbold{p}}]^2$, $t =
 -2\mbold{p}^2(1-{\rm cos}\,\theta)$ and $\theta$ is the scattering angle in
 the c.m. system.
 
 The real part of the partial wave amplitudes near threshold is of the form,
 \begin{equation}
   {\rm Re}\,t^I_l(s) =
   \frac{\sqrt{s}}{2}\mbold{p}^{2l}\left(a^I_l+b^I_l \mbold{p}^2+\Order{(\mbold{p}^4)}\right),
 \end{equation}
 where $a^I_l$ denote the scattering lengths and $b^I_l$ the effective ranges.

What follows, are the $\pi\pi$ scattering processes:
\begin{equation}
  \pi^- \pi^+ \rightarrow \pi^- \pi^+, \quad \pi^- \pi^+ \rightarrow \pi^0 \pi^0,
\end{equation}
with
\begin{eqnarray}
  T^{\pm;\pm}_\pi &=& \frac{1}{6}\left[ T^2_\pi(s, t)+3T^1_\pi(s,t)+2T^0_\pi(s,t)\right],\nn\\
T^{00;\pm}_\pi &=& \frac{1}{3}\left[T^2_\pi(s,t)-T^0_\pi(s,t)\right].
\end{eqnarray}
The decomposition into partial waves yields
\begin{equation}
  T^I_\pi(s,t) = 32\pi\sum_{l=0}^\infty (2l+1)P_l({\rm cos}\,\theta)t^I_{\pi, l}(s),
\end{equation}
where $s = 4(\Mpic^2+\mbold{p}^2)$. At
threshold, the partial wave amplitudes take the form
\begin{equation}
  {\rm Re}t_{\pi, l}^I(s) = \mbold{p}^{2l}\left[a_l^I+\mbold{p}^2b_l^I+\Order{(\mbold{p}^4)}\right].
\end{equation}
\setcounter{equation}{0}
\section{Schwinger's Green function}
\label{appendix: green function}
The Schwinger Green function fulfills
\begin{eqnarray}
 \left[E-\frac{\mbold{q}^2}{2\mu_+}\right]{}_+\!(\mbold{q}\!\mid\mbold{G}_{\sss
  \rm C}(z)\mid\!\mbold{p})_+ +e^2\int \dnu(k)\frac{1}{(\mbold{k}-\mbold{q})^2}{}_+\!(\mbold{k}\!\mid\mbold{G}_{\sss
  \rm C}(z)\mid\!\mbold{p})_+ =\nn\\
(2\pi)^3\delta^3(\mbold{q}-\mbold{p}),
\end{eqnarray}
where $E = z-\Sigma_+$. The explicit expression is given by\footnote{To simplify the notation, we
  omit the positive imaginary part in E.}
\begin{eqnarray}
  {}_+\!(\mbold{q}\!\mid\mbold{G}_{\sss \rm C}(z)\mid\!\mbold{p})_+ &=&
  \frac{(2\pi)^3\delta^3(\mbold{q}-\mbold{p})}{E-\frac{\mbold{q}^2}{2\mu_+}}-\frac{1}{E-\frac{\mbold{q}^2}{2\mu_+}}\frac{4\pi\alpha}{(\mbold{q}-\mbold{p})^2}\frac{1}{E-\frac{\mbold{p}^2}{2\mu_+}}\nn\\
&&-\frac{1}{E-\frac{\mbold{q}^2}{2\mu_+}}4\pi\alpha\eta I(E,\mbold{q},\mbold{p})\frac{1}{E-\frac{\mbold{p}^2}{2\mu_+}},
\end{eqnarray}
with $\eta = \alpha/2(-E/(2\mu_+))^{-1/2}$. The function 
\begin{equation}
  I(E,\mbold{q},\mbold{p}) = \int_0^1 d\mbold{\rho} \,\rho^{-\eta}\left[(\mbold{q}-\mbold{p})^2\rho+\frac{\eta^2}{\alpha^2}(1-\rho^2)(E-\frac{\mbold{q}^2}{2\mu_+})(E-\frac{\mbold{p}^2}{2\mu_+})\right]^{-1},
\end{equation}
contains poles at $\eta = 1,2,\dots$ or, equivalently, at $z = E_n$. The integral
\begin{eqnarray}
\langle\mbold{\bar{g}}^n_{\sss \rm C}(E_n)\rangle &=& \int
  \frac{d^d\mbold{p}}{(2\pi)^d}
  \frac{d^d\mbold{q}}{(2\pi)^d}{}_+\!(\mbold{q}\!\mid\mbold{\bar{G}}^n_{\sss \rm
  C}(E_n)\mid\!\mbold{p})_+,\nn\\
  &=&
  \frac{\alpha\mu_+^2}{\pi}\left\{\psi(n)-\psi(1)-\frac{1}{n}+\frac{1}{2}\left[\Lambda(\mu)-1+2{\rm
  ln}\frac{\alpha}{n}+{\rm ln}\frac{4\mu_+^2}{\mu^2}\right]\right\},\nn\\
\end{eqnarray}
with $\psi(n)=\Gamma'(n)/\Gamma(n)$ develops an ultraviolet singularity as $d\rightarrow3$,
\begin{equation}
  \Lambda(\mu) =(\mu^2)^{d-3}\left\{\frac{1}{d-3}-\Gamma'(1)-{\rm
    ln}4\pi\right\}.
\label{Lambda}
\end{equation}
\setcounter{equation}{0}
\section{Non-relativistic integrals}
\label{appendix: integrals}
What follows is a list of the non-relativistic integrals used to calculate the
$\pi K$ scattering amplitudes in Section \ref{section: matching} as well as in
the evaluation of the decay width (Sec. \ref{section: resolvents}) and energy
shifts (Secs. \ref{section: resolvents} and \ref{section: emEnergy}). Whenever
necessary, the integrations are worked out in $D\neq4$ space-time
dimensions to take care of possible ultraviolet or infrared divergences.  

The non-relativistic propagator of the heavy fields reads
\begin{equation}
  G_{{\rm NR}, h}(x) \doteq i\langle 0 | T \bar{h}(x) \bar{h}^\dagger(0)|0\rangle =
  \int \frac{d^D
  p}{(2\pi)^D}e^{-ipx}\frac{1}{M_h+\frac{\mbold{p}^2}{2M_h}-p^0-i\epsilon} ,
\label{propagator}
\end{equation}
where $\bar{h}$ stands for the free fields, with $h = \pi^-, \pi^0, K^+, K^0$
and $M_h$ denotes the corresponding mass. The tadpole diagrams $G_{{\rm NR},
  h}(0)$ vanishes in dimensional regularization. This can be seen, by
performing the integration over the zero component of the loop momentum explicitly. The
remaining integral is scaleless and therefore zero in dimensional regularization.

At $\alpha = 0$, the elementary loop function to calculate a diagram with any
number of bubbles is given by
\begin{equation}
  J_i(P^0) = \frac{1}{i}\int \frac{d^D
  l}{(2\pi)^D}\frac{1}{M_{\pi^i}+\frac{\mbold{l}^2}{2M_{\pi^i}}-l^0}\frac{1}{M_{\sss
  K^i}+\frac{\mbold{l}^2}{2M_{\sss K^i}}-P^0+l^0},
\end{equation}
where $i = +,0$. After the integration over the zero component of the loop
momentum, we arrive at
\begin{eqnarray}
  J_i(P^0) &=& \int \frac{d^d
  \mbold{l}}{(2\pi)^d}\frac{1}{\Sigma_i+\frac{\mbold{l}^2}{2\mu_i}-P^0}.
\label{Ji}
\end{eqnarray}
The function is analytic in the complex $P^0$ plane, except for a cut on the
real axis for $P^0 \ge \Sigma_i$. For $P^0 \ge \Sigma_i$ and $d\neq3$, we get
\begin{eqnarray}
  J_i(P^0) &=& \frac{i\mu_i^{3/2}}{\sqrt{2}
  \pi}\sqrt{P^0-\Sigma_i}\bigg[1+\frac{d-3}{2}\nn\\
&&
\times\left(-i\pi-2-\Gamma'(1)+{\rm ln}\, 4\pi+{\rm
  ln}\,2\mu_i(P^0-\Sigma_i)\right)\bigg].
\label{Jia}
\end{eqnarray}
 To obtain the contribution to the scattering amplitude, we need to evaluate this function at $P^0 =
\w{{\pi^+}}{\mbold{p}}+\w{{K^+}}{\mbold{p}}$. At threshold, the
integral $J_+(P^0)$ is of order $\mbold{p}$, while $J_0(P^0)$ is given by
\begin{equation}
  J_0(\Sigma_+) = \frac{i \mu_0^{3/2}}{\sqrt{2}\pi}\sqrt{\Sigma_+-\Sigma_0}.
\end{equation}
We now include the Coulomb interaction. The self-energy diagram with one
virtual Coulomb photon vanishes, because the integration contour over the zero
momentum of the photon can be closed in the upper half-plane, where there is no
singularity. Next, we evaluate the Coulomb vertex
function $V_{\sss \rm C}$
and the two--loop integral $B_{\sss \rm C}$. The Coulomb vertex function, see
for example Figure \ref{fig: Coulomb}(a), is given by  
\begin{eqnarray}
  V_{\sss \rm C}(\mbold{p}, P^0)& =& -e^2\frac{1}{i}\int
  \frac{d^Dl}{(2\pi)^D}\frac{1}{|\mbold{p}-\mbold{l}|^2}
\frac{1}{\Mpic+\frac{\mbold{l}^2}{2\Mpic}-l^0}\nn\\
&&\times\frac{1}{\MKc+\frac{\mbold{l}^2}{2\MKc}-P^0+l^0}.
\end{eqnarray}
After integrating over the zero component of the loop
momentum, the function amounts to
\begin{equation}
  V_{\sss \rm C}(\mbold{p}, P^0) = e^2\int
  \frac{d^d\mbold{l}}{(2\pi)^d}\frac{1}{\mid\!\mbold{p}-\mbold{l}\!\mid^2}\frac{1}{P^0-\Sigma_+-\frac{\mbold{l}^2}{2\mu_+}}.
\end{equation}
The contribution to the scattering amplitude is obtained for $P^0=
\w{{\pi^+}}{\mbold{p}}+\w{{K^+}}{\mbold{p}}$. We expand the function
around threshold which leads to
\begin{equation}
  V_{\sss \rm C}(\mbold{p},P^0_{\rm thr}) = -\frac{\pi\alpha\mu_+}{2\mid\!\mbold{p}\!\mid}-i\alpha\theta_c+\Order{(d-3)},
\end{equation}
where $P^0_{\rm thr} = \Sigma_++\frac{\mbold{p}^2}{2\mu_+}$ and the
infrared-divergent Coulomb phase $\theta_c$ is specified in
Eq. (\ref{Coulombphase}). The two--loop Coulomb photon exchange diagram depicted in Figure \ref{fig: Coulomb}(c) reads
\begin{eqnarray}
  B_{\sss \rm C}(P^0) &=& -e^2 \int
  \frac{d^Dl_1}{(2\pi)^D}\frac{d^Dl_2}{(2\pi)^D}\frac{1}{|\mbold{l}_1-\mbold{l}_2|^2}\nn\\
&&\times\frac{1}{\Mpic+\frac{\mbold{l}_1^2}{2\Mpic}-l_1^0}
  \frac{1}{\MKc+\frac{\mbold{l}_1^2}{2\MKc}-P^0+l_1^0}\nn\\
&&\times\frac{1}{\Mpic+\frac{\mbold{l}_2^2}{2\Mpic}-l_2^0}
\frac{1}{\MKc+\frac{\mbold{l}_2^2}{2\MKc}-P^0+l_2^0}.
\end{eqnarray}
Performing the integrations over the zero components of the loop momenta $l_1$
and $l_2$, we get
\begin{equation}
  B_{\sss \rm C}(P^0) = e^2 \int \frac{d^d\mbold{l}_1}{(2\pi)^d}\frac{d^d\mbold{l}_2}{(2\pi)^d}\frac{1}{\mid\!\mbold{l}_1-\mbold{l}_2\!\mid^2}\frac{1}{P^0-\Sigma_+-\frac{\mbold{l}_1^2}{2\mu_+}}\frac{1}{P^0-\Sigma_+-\frac{\mbold{l}_2^2}{2\mu_+}}.
\end{equation}
 The expansion in the vicinity of the threshold amounts to 
\begin{equation}
  B_{\sss \rm
  C}(P^0_{\rm thr})=-\frac{\alpha\mu_+^2}{2\pi}\left[\Lambda(\mu)+2{\rm ln}\frac{2\mid\!\mbold{p}\!\mid}{\mu}-1-i\pi\right]+\Order{(d-3)}.
\end{equation}
The ultraviolet pole term $\Lambda(\mu)$ is given in Eq. (\ref{Lambda}). 

In the presence of transverse photons, the non-relativistic integrals have to
be treated properly in order to avoid loop momenta coming from the hard
scale. Otherwise, these loop corrections lead to a breakdown of the
non-relativistic counting scheme. We use the threshold expansion
\cite{threxp1,threxp2} to disentangle the hard scale (given by the meson
masses) from the soft scales. We illustrate the procedure for the two-point
function of the charged pions and kaons at order $e^2$,
\begin{equation}
  i\int dx e^{ipx}\langle 0|T h_\pm(x)h^\dagger_\pm(0)|0\rangle
  =\frac{1}{M_++\frac{\mbold{p}^2}{2M_+}-p^0-\Sigma_h(p^0, \mbold{p})},
\end{equation}
where $h=\pi,K$ and $M_+$ denotes the corresponding mass. The
self-energy $\Sigma_h$ is depicted in Figure \ref{fig: transverse}(a) (upper
line). In $D\neq 4$ space-time dimensions, we have
\begin{equation}
  \Sigma_{h}(p^0, \mbold{p}) = \frac{e^2}{M_+^2}\frac{1}{i}\int \frac{d^D k}{(2\pi)^D}\frac{
  \mbold{p}^2-(\mbold{p}\cdot\mbold{k})^2/\mbold{k}^2}{-k^2(M_++\frac{(\mbold{p}-\mbold{k})^2}{2M_+}-p^0+k^0)}+\Order{(e^4)}.
\label{selfenergyD}
\end{equation}
After integrating over the zero component of the loop momentum, we arrive
at 
\begin{equation}
  \Sigma_{h}(p^0, \mbold{p})=\frac{e^2}{M_+^2}\int \frac{d^d
  k}{(2\pi)^d}\frac{1}{2|\mbold{k}|}\frac{\mbold{p}^2-(\mbold{p}\cdot\mbold{k})^2/\mbold{k}^2}{\Omega+\frac{\mbold{k}^2}{2M_+}-\frac{\mbold{p}\cdot\mbold{k}}{M_+}+|\mbold{k}|}+\Order{(e^4)},
\label{selfenergy}
\end{equation}
where $\Omega = M_+ +\mbold{p}^2/(2M_+)-p^0$. The threshold expansion amounts
 to expanding the integrand in Eq. (\ref{selfenergy}) in the small parameter $v$, according to the
 counting\footnote{Instead of first performing the integration over the zero component of the
 photon field, we may apply the threshold expansion directly to
 Eq. (\ref{selfenergyD}), with $k^0=\Order{(v^2)}$.
},
\begin{equation}
  \mbold{k} = \Order{(v^2)}, \quad \mbold{p} =  \Order{(v)}, \quad \Omega = \Order{(v^2)}.
\end{equation}
The threshold expanded self-energy
\begin{equation}
  \hat{\Sigma}_h(\Omega, \mbold{p}) = \frac{e^2}{2M_+^2}\mbold{p}^2\Omega^{d-2}\frac{\Gamma(d)\Gamma(2-d)}{(4\pi)^{d/2}\Gamma(1+\frac{d}{2})}+\Order{(e^4)},
\end{equation}
contains an ultraviolet divergence as $d\rightarrow 3$,
\begin{eqnarray}
  \hat{\Sigma}_h(\Omega, \mbold{p}) &=&
  \frac{e^2}{6\pi^2M_+^2}\mbold{p}^2\Omega\left(L(\mu)+{\rm
  ln}\frac{2\Omega}{\mu}-\frac{1}{3}\right)+\Order{(e^4, d-3)},\nn\\
L(\mu) &=& \mu^{d-3}\left[\frac{1}{d-3}-\frac{1}{2}\left(\Gamma'(1)+{\rm ln}4\pi+1\right)\right].
\label{thrselfenergy}
\end{eqnarray}
\setcounter{equation}{0}
\section{Unitarity condition: Evaluation of the width}
\label{appendix: unitarity condition}
The unitarity condition
in Eq. (\ref{unitarity}) renders the evaluation of the S-wave decay width at
next-to-leading order straightforward. It can be seen from Eqs. (\ref{wavefunction}) and (\ref{phasespace}) that
in order to evaluate the width at $\Order{(\delta^{9/2})}$, it suffices to
  calculate the matrix element $_+\!( \mbold{p}\!\mid
\mathbold{\bar{\tau}}(E_n)\mid\!\mbold{p}_3)_0$ at order $\delta$.
Here, the following term occurs 
 \begin{eqnarray}
   \Gamma_{n0, \pi^0 K^0} &=&
   -C_2^2\frac{\Mpin^3+\MKn^3}{8\Mpin^3\MKn^3}|\Psi_{n0}(\mbold{x}=0)|^2\Int
   \dnu(\mbold{p}_3)2\pi\delta\Big(E_n-\Sigma_0-\frac{\mbold{p}_3^2}{2\mu_0}\Big)\nn\\
 &&
   \times\mbold{p}_3^4\left[\frac{1}{z-\Sigma_0-\frac{\mbold{p}_3^2}{2\mu_0}}+\frac{1}{\bar{z}-\Sigma_0-\frac{\mbold{p}_3^2}{2\mu_0}}\right]\Bigg|_{z\rightarrow
   E_n+i \epsilon}+\cdots,
 \end{eqnarray}
which is generated by the matrix element $_+\!(\mbold{p}\!\mid\!\mbold{H}_{\sss \rm S}\mbold{G}_{\sss \rm C}(z)\mbold{H}_{\sss \rm
  D}\!\mid\!\mbold{p}_3)_0$ and its hermitian conjugate. This contribution
can be calculated by the use of
\begin{eqnarray}
  -2\pi i\delta\left(E_n-\Sigma_0-\frac{\mbold{p}_3^2}{\mu_0}\right) =
  \frac{1}{z-\Sigma_0-\frac{\mbold{p}_3^2}{\mu_0}}-\frac{1}{\bar{z}-\Sigma_0-\frac{\mbold{p}_3^2}{\mu_0}}\Bigg|_{z\rightarrow
  E_n+i \epsilon},
\end{eqnarray}
and the result for the decay width at order $\delta^{9/2}$ agrees with
Eq. (\ref{decaywidth}).
\end{appendix}

\end{document}